\documentclass[amsmath,superscriptaddress,nofootinbib,showpacs,pre]{revtex4}
\pdfoutput=1
\usepackage{graphicx}
\usepackage{amsmath}
\usepackage{epsfig}
\usepackage{graphics}
\usepackage{color}
\usepackage{latexsym}
\usepackage{etoolbox}
\usepackage{amsfonts}
\usepackage{amssymb}
\usepackage{color}  
\usepackage{plain}   

\begin{document}

\title{Feedback cooling, measurement errors, and entropy production}
\author{T. Munakata}
\affiliation{Department of Applied Mathematics and Physics, 
Graduate School of Informatics, Kyoto University, Kyoto 606-8501, Japan
} 
\email{tmmm3rtk@hb.tp1.jp}
\author{M.L. Rosinberg}
\affiliation{Laboratoire de Physique Th\'eorique de la Mati\`ere Condens\'ee, Universit\'e Pierre et Marie Curie\\ 4 place Jussieu, 75252 Paris Cedex 05, France}
\email{mlr@lptmc.jussieu.fr}

%\date{\today}
\begin{abstract}
The efficiency of a feedback mechanism depends on the precision of the measurement outcomes obtained from the controlled system. Accordingly, measurement errors  affect the entropy production in the system. We explore this issue in the context of active feedback cooling by modeling a typical cold damping setup as a harmonic oscillator in contact with a heat reservoir and submitted to a velocity-dependent feedback force that reduces the random motion. We consider two models that distinguish whether the sensor continuously measures the position of the resonator or directly its velocity (in practice, an electric current).   Adopting the standpoint of the controlled system, we identify the `entropy pumping'  contribution that describes the  entropy reduction due to the  feedback control and that modifies  the second law of thermodynamics. We also assign a relaxation dynamics to the feedback mechanism and compare the apparent entropy production in the system and the heat bath (under the influence of  the controller) to the total entropy production in the super-system that includes the controller. In this context, entropy pumping reflects the existence of hidden degrees of freedom and the apparent entropy production satisfies fluctuation theorems associated to an effective Langevin dynamics.
\end{abstract} 
\pacs{05.40.-a, 05.10.Gg, 05.70.Ln}
\maketitle
\section{Introduction}

Active feedback cooling is a well-established technique which is used to reduce the effective noise temperature  of mechanical oscillators well below their operating temperature.  It is now used in a wide variety of nano-electromechanical systems, and is a key ingredient for measuring force and mass with very high sensitivity, for limiting thermal noise in gravitational waves detectors, and for reaching the quantum regime of mechanical motion\cite{PZ2012}.  
A commonly used procedure named  {\it cold damping} consists in measuring the resonator displacement in real time and then applying through a feedback loop a velocity-proportional external force that increases the damping rate. As a result, the Brownian motion of the resonator (for instance, the mirror of an interferometric detector\cite{PCBH2000} or the cantilever of an AFM\cite{PDMR2007}) is reduced. Ultimately, the feedback cooling efficiency (that is the minimum achievable temperature) is  bounded by the noise of the detector, that is by the errors in the measurements. 

Such a feedback loop thus plays the role of an external agent that detects the microscopic state of the system and then acts to modify its  dynamical evolution. It is therefore natural to wonder how the information acquired through the measurement modifies the thermodynamics of the system,  in particular the entropy balance equation and the second law. This issue, which is at the crossroad of information theory and nonequilibrium statistical mechanics, has attracted much attention recently\cite{SU2008,CF2009,P2010,FS2010,HV2010,T2010,HP2011,IS2011,CF2012,AS2012,SU2012,SU2012b}, in relation with significant advances in single-molecule manipulations and  new fundamental developments in the stochastic thermodynamics of small systems\cite{Sbook2010,J2011,Sreview2012}. Within this framework, nonequilibrium  relations such as Jarzynski equality\cite{J1997} and fluctuation theorems (FT)\cite{GLKD2012} have been generalized to systems under {\it discrete} feedback control by taking into account the mutual information between the state of the system and the measurement outcome. Mutual information  quantifies the entropy reduction due to the interaction with the external agent (hereafter also called the {\it controller}) and provides a lower bound to the entropy production (EP) in the system. Measurement errors decrease mutual information, and thus limit the entropy reduction and the efficiency of the feedback control. 
 
These results, however, are not directly applicable to cold damping. First,  measurements and actuation are performed {\it continuously}  in this process, i.e., repeated with a period shorter than the characteristic time scales of the system dynamics. In practice, the motion of the feedback-cooled resonator in the vicinity of a resonant frequency is faithfully described by an underdamped Langevin dynamics. Second, the feedback controller is not a genuine Maxwell's demon that only exchanges information with the system\footnote{Throughout this paper, the term `feedback controller' denotes both the sensor that measures the state of the system and the actuator that modifies its dynamics.}. Indeed, the feedback force performs work on the system and the energetics (i.e. the first law) is thus modified. One expects the formulation of the second law to be modified as well, even for error-free measurements (in such a situation, mutual information diverges since observables are continuous\cite{IS2011,SU2012b}). This issue was first explored in \cite{KQ2004} where it was shown that entropy is indeed continuously pumped out of the system by the external agent in the nonequilibrium steady state (NESS). This can be ascribed to the contraction of momentum phase space due to the additional damping. This {\it entropy pumping} modifies the second law, Jarzynski equality, and fluctuation theorems\cite{KQ2007}. The fact that the feedback force is velocity-dependent makes the formulation of the detailed FT somewhat peculiar, as stressed in a previous work\cite{MR2012}.

The goal of this paper is to include the effect of  measurement errors in this description.  How is the entropy pumping in the NESS affected by the detector noise ? Although the influence of noise on the EP in a cold damping process has been explored in a recent work\cite{K2012}, this question has not been addressed\cite{GBPC2010}. Indeed, the quantity studied in \cite{K2012} is the  EP in the ``super-system" that includes the Brownian entity (i.e. the resonator), the heat bath, and the feedback controller. The contribution of the controller to the EP is  interpreted as resulting from the measurement process and is obtained by taking an appropriate continuous limit of a discrete series of noisy measurements\footnote{The relationship between a {\it single} noisy measurement of the velocity and the violation of the fluctuation dissipation theorem in a cold damping process is investigated in \cite{IS2011}.}. In the super-system, the usual second law is obeyed and entropy pumping plays no role. Furthermore, the  EP diverges in the limit of error-free measurements and the results of \cite{KQ2004,KQ2007} are thus not recovered. This clearly shows that the level of description of the system is different. To generalize the analysis of \cite{KQ2004,KQ2007,MR2012}, one must instead consider the EP from the viewpoint of the controlled system\footnote{In this respect, the use of the term ``super-system'' in our previous work\cite{MR2012} was inappropriate.}:  entropy pumping (like mutual information) then describes the entropy reduction due to the interaction of the system with the controller\cite{CF2012}.  This is also  the viewpoint adopted in the information-theoretic approach of control systems\cite{TL2000}. At this level of description, one is not interested in knowing the actual energy dissipation inside the controller nor in evaluating the  ``thermodynamic" cost of measurement\cite{GK2011} (see also \cite{SU2012b} and references therein)\footnote{More generally, one does not gain much insight into the physics of the problem by investigating the EP in the super-system unless one has a precise description of the actual mechanisms that take place inside the controller. The case of genuine Maxwell's demons is different and the problem can be investigated on a case-by-case basis (see e.g. \cite{SSBE2012}).}.

This important distinction between the two levels of description becomes even clearer when the controller has its proper dynamics. The super-system is then characterized by several (slow) degrees of freedom and adopting the standpoint of the controlled system means that the degree(s) of freedom of the controller is (are) projected out. 
In other words, the super-system is coarse-grained, which in turn changes the definition of the EP. This issue of coarse graining has also attracted much attention recently\cite{RJ2007,KPVdB2007,GPVdB2008,VdBE2010,RP2012,E2012,CPV2012,FS2012,NK2012}, including on the experimental side\cite{MLBBS2012}. As a rule, one expects that an incomplete description of the dynamics, due to the existence of hidden degrees of freedom or to a low resolution measuring apparatus, results in an underestimation of the EP in the full system. 

In this work, we shall indeed assign a relaxation dynamics to the controller. Physically, this may account for the fact that the feedback circuit cannot follow instantaneously the dynamics of the system when the resonator frequency is high (say above $1$MHz). This makes the feedback control non-Markovian and degrades the cooling performance\cite{PZ2012}.  On the other hand, this procedure allows us to study the EP in a typical cold damping setup in which the position of the resonator (for instance a cantilever\cite{PDMR2007,L2000,JTCC2007,WOY2010}, a suspended mirror in a cavity\cite{PCBH2000}, or an optically trapped microsphere\cite{LKR2011}) is continuously monitored.  In this case, the velocity-proportional feedback force also includes the derivative of the detector noise\footnote{See e.g. Fig. 22 in \cite{PZ2012} for a schematic diagram of a cold damping setup.}, which is obviously problematic when the noise spectrum is flat (i.e. the noise is white), as is usually assumed (see e.g. {\cite{PCBH2000,PDMR2007,WOY2010}). In this case, the relaxation time plays the role of a cut-off that regularizes divergent quantities. On the other hand, no regularization is needed when the measured observable is directly the velocity (in practice, the  current in a RLC electric circuit), a situation that corresponds to the electronic feedback cooling schemes described in \cite{V2008,B2009,VBMF2010}. The theoretical analysis of this second setup is  simpler and the Markovian limit of the feedback control is well defined (this is actually the case considered in \cite{IS2011,K2012}).
 
The rest of the paper is organized as follows. In section 2, we specify the two models (hereafter called P for position and V for velocity) that are investigated and that correspond to the two experimental situations that have just been described. In section 3, we review the results of \cite{KQ2004,KQ2007,MR2012} in order to make the whole story of the paper self-explanatory and coherent. In section 4, measurement errors are included and in section 4A model V is first studied in the Markovian limit. This is the simplest case for which the expression of entropy pumping can be generalized. The full versions of models V and P are then successively studied in section 4B and 4C. We summarize the main points and conclude in section 5.  Some additional (but important) calculations are detailed in three Appendices. 

\section{Models and equations of motion}

The physical systems studied in this paper are described by the underdamped one-dimensional Langevin equation
\begin{align}
\label{EqL1}
m\ddot x+\gamma \dot x+kx=F_{\textnormal{th}}(t) +F_{\textnormal{fb}}(t)
\end{align}
where $x(t)$ denotes the position of the  resonator as a function of time, $m$ is an effective mass, $k$  is a spring constant, and $\gamma$ is a linear damping. As usual, the resonator dynamics can  be also described in terms of the angular resonance frequency $\omega_0=\sqrt{k/m}$ and the intrinsic quality factor $Q_0=\omega_0\tau_0=\sqrt{mk}/\gamma$, where $\tau_0=m/\gamma$ is  the viscous relaxation time. Eq. (\ref{EqL1}) correctly describes the small displacements of nanomechanical systems around the resonance frequency of a normal mode\cite{PZ2012}. Alternatively, it may also describe a RLC electrical circuit: $x(t)$  then represents the charge of the capacitor whereas the velocity $v(t)\equiv \dot x(t)$ is the current through the inductor (with the resistor $R$, inductor $L$, and capacitor $C$ such that $\gamma =R/L$, $\omega_0^2= 1/LC$ and $m = L$). One may also simply regard Eq. (\ref{EqL1})  as modeling the dynamics of a Brownian particle confined by a harmonic potential. 

$F_{\textnormal{th}}(t)=\sqrt{2\gamma T}\: \xi(t)$ represents the thermal random force generated by the surrounding medium at temperature  $T$, and $\xi(t)$ is a Gaussian white noise with zero mean and correlation $<\xi(t)\xi(t')>=\delta (t-t')$ (for notational simplicity,  Boltzmann's constant is adsorbed in the temperature throughout this paper). $F_{\textnormal{fb}}(t)$ is the velocity-proportional control force which is applied to the resonator via the feedback loop. We consider the following two models:

(i) model  V, in which the velocity $v(t)$ is continuously measured and the output signal of the detector is $v'(t)=v(t)+v_n(t)$, where $v_n(t)$ is the measurement noise. The feedback force is then  given by 
\begin{align}
\label{EqV}
F_{\textnormal{fb}}(y)=-\gamma' y(t)
\end{align}
where $\gamma'=g\gamma$ ($g$ is the variable gain of the feedback loop), and $y(t)$ obeys the differential equation
\begin{align}
\label{EqyVV}
\dot y=-\frac{1}{\tau} [y -v'] 
\end{align}
where $\tau$ is the relaxation time of the feedback circuit.

(ii) model P, in which the  displacement $x(t)$ is the observable and the output signal of the detector is $x'(t)=x(t)+x_n(t)$, where $x_n(t)$ is the measurement noise. In this case
\begin{align}
\label{EqMP}
F_{\textnormal{fb}}(\dot y)=-\gamma' \dot y(t)
\end{align}
where 
\begin{align}
\label{EqyPP}
\dot y=-\frac{1}{\tau} [y -x'] \ .
\end{align}

Since it is generally observed that the power spectral density of the measurement noise is flat in the frequency band of interest, we assume that $v_n(t)$ and $x_n(t)$ are delta-correlated in time:
\begin{align}
v_n(t)&=\sqrt{S_{v_n}} \:\eta (t)\nonumber\\
x_n(t)&=\sqrt{S_{x_n}}\: \eta (t)
\end{align}
where $\eta(t)$ is a Gaussian white noise with zero mean and correlation $<\eta(t)\eta(t')>=\delta (t-t')$. We moreover assume that the two noises $\eta(t)$ and $\xi(t)$ are uncorrelated and that the spectral densities $S_{v_n}$ and $S_{x_n}$ do not depend on the gain $g$, which is a reasonable approximation\footnote{In experiments,  the detector noise is usually obtained by fitting the measured spectral density with a theoretical expression such as  Eq. (\ref{EqPSDVb}) or Eq.  (\ref{EqPSDPb}) (with $\tau=0$). It is found that the noise parameters do not depend significantly on the feedback gain $g$ (see e.g. \cite{PDMR2007} in the case of a mechanical resonator and \cite{VBMF2010} in the case of an electrical resonator).}. 

Model V is thus defined by the two coupled linear Langevin equations
\begin{subequations}
\label{EqmodelV}
\begin{align}
\label{EqmodelVa}
m\ddot x+\gamma \dot x+kx +\gamma' y &=\sqrt{2\gamma T}\: \xi(t)\\
\label{EqmodelVb}
\tau \dot y+y -\dot x&=\sqrt{S_{v_n}}\: \eta (t) \ ,
\end{align}
\end{subequations}
whereas model P is defined by 
\begin{subequations}
\label{EqmodelP0}
\begin{align}
\label{EqmodelP0a}
m\ddot x+\gamma \dot x+kx +\gamma'\dot  y &=\sqrt{2\gamma T}\: \xi(t)\\
\label{EqmodelP0b}
\tau \dot y+y -x&= \sqrt{S_{x_n}}\: \eta (t)\ ,
\end{align}
\end{subequations}
or better 
\begin{subequations}
\label{EqmodelP}
\begin{align}
\label{EqmodelPa}
m\ddot x+\gamma \dot x+kx+\frac{\gamma'}{\tau}(x-y)&=\sqrt{2\gamma T}\: \xi(t) -\frac{\gamma'}{\tau} \sqrt{S_{x_n}}\: \eta (t) \\
\label{EqmodelPb}
\tau \dot y+y -x&=\sqrt{S_{x_n}}\: \eta (t)\ ,
\end{align} 
\end{subequations}
which is the form under which the model can be numerically studied.

It is worth noting that the  physical processes described by the above equations are Markovian if both $x(t)$ and $y(t)$ are observed, whereas  the effective dynamics of $x(t)$ obtained by solving Eqs. (\ref{EqmodelVb}) or (\ref{EqmodelPb}) for $y(t)$ and inserting the result into (\ref{EqmodelVa}) or (\ref{EqmodelPa}) is non-Markovian\footnote{The influence of memory terms on the behavior of feedback-controlled harmonic oscillators is also considered in \cite{GRBC2009} in reference to feedback-cooled electromechanical oscillators, such as the gravitational wave detector AURIGA\cite{V2008,B2009}. In that study, however, the detector noise is not taken into account and the stochastic thermodynamics of the system is not investigated.}.
This transformation is discussed in detail in \cite{VBPV2009, PV2009} and more recently in \cite{CPV2012} for a system of  coupled linear Langevin equations quite similar to  Eqs. (\ref{EqmodelV}). These equations were originally regarded as modeling the irreversible dynamics of a massive tracer in a granular fluid\cite{SVCP2010}, which is a quite different physical situation from the one considered here. However, the discussion in \cite{CPV2012} about the influence of cross-correlations among different degrees of freedom on the entropy production is relevant to the present work.%Moreover, in the so-called ``analog" protocol, which leads to a Langevin equation similar to the one considered in model V, the memory kernel describes a low-pass filter with a cut-off frequency much smaller than the resonance frequency. This would correspond to a large value of $\tau$ in model V.

A Markovian description is of course recovered in the limit $\tau \rightarrow 0$ as $y(t)\rightarrow v'(t)$ and $y(t) \rightarrow x'(t) $ in model V and P, respectively (note that the measurement itself is always Markovian since the measurement outcomes $v'$ and $x'$ do not depend on the state of the system at previous times).  The motion of the resonator is then simply described by
\begin{align}
\label{EqL2V}
m\ddot x+(\gamma +\gamma')\dot x+kx=\sqrt{2\gamma T}\xi(t) -\gamma' v_n(t)
\end{align}
in model V, and by
\begin{align}
\label{EqL2P}
m\ddot x+(\gamma +\gamma')\dot x +kx=\sqrt{2\gamma T}\xi(t) -\gamma' \dot x_n(t)
\end{align}
in model P. Eq. (\ref{EqL2V}) with $k=0$ is the equation studied in \cite{IS2011,K2012}. It also corresponds to a simplified version of the model considered in \cite{VBMF2010} that describes the feedback cooling of a macroscopic electromechanical oscillator. On the other hand, Eq. (\ref{EqL2P}) is typically used to describe the feedback cooling of a cantilever\cite{PDMR2007}. It is clear, however, that this  equation is ill-defined mathematically if $x_n(t)$ is a white noise. Therefore, as we already pointed out, the introduction of a finite relaxation time $\tau$ may also be regarded as a way to circumvent this problem without having to describe the  frequency dependence of the measurement noise, which is generally unknown. In practice, for the cooling to be efficient, $\tau$ must be much smaller than $m/(\gamma+\gamma')$, the effective viscous relaxation time of the feedback-controlled oscillator, so that $y(t)$ (in model V) and $\dot y(t)$ (in model P) follow $v(t)$ fast enough. In other words, the dynamics of the controller must be much faster than the dynamics of the system, otherwise the information about the instantaneous velocity is lost. 

\section{Entropy production with error-free measurements: a reminder}

The entropy production in a cold damping setup in the absence of measurement errors was first investigated in \cite{KQ2004,KQ2007} and revisited by us in a previous work\cite{MR2012}. The feedback force is given by
\begin{align}
F_{\textnormal{fb}}(v)=-\gamma' v(t) \ 
\end{align}
and the system is thus described by the  Langevin equation 
\begin{align}
\label{EqLKQ}
m\ddot x+(\gamma+ \gamma')\dot x +kx=\sqrt{2\gamma T}\xi(t) 
\end{align}
which is also obtained from Eqs. (\ref{EqL2V}) and (\ref{EqL2P}) for $v_n(t)=x_n(t)=0$.  Accordingly, the probability distribution function $p_t(x,v)$ at the ensemble level satisfies  the Fokker-Planck (FP) equation
\begin{align}
\partial_t  p_t(x,v)=-\partial_x [vp_t(x,v)] +\frac{1}{m}\partial_v\left[[kx+(\gamma+\gamma')v]p_t(x,v) +\frac{\gamma T}{m} \partial_v p_t(x,v)\right] 
\end{align}
 which is conveniently rewritten as 
\begin{align}
\label{EqFPKQ}
\partial_t  p_t(x,v)=-\partial_x [vp_t(x,v)] -\frac{1}{m}\partial_v[-(kx+\gamma' v) p_t(x,v)+ J_t(x,v)] \ ,
\end{align}
where 
\begin{align}
\label{EqJ0}
 J_t(x,v)=-\gamma [ v+\frac{T}{m} \partial_v \ln  p_t(x,v)] p_t(x,v)
\end{align}
is a probability current (in what follows, time is often put as an index for better readability). Although one can directly derive the entropy balance equation at the ensemble level\cite{KQ2004}, it is helpful to consider the various thermodynamic quantities at the level of an individual stochastic trajectory, as done in \cite{KQ2007,MR2012}, in order to better understand the origin of entropy pumping. The ensemble average is then taken in a second stage. 

Let $\{x_s\}_{s\in [0,t]}$ denote a trajectory generated by Eq. (\ref{EqLKQ}) during the time interval $0\le s\le t$ with an initial state drawn from some probability distribution $p_0(x,v)$. Within the framework of stochastic energetics\cite{Sbook2010},  the energy balance equation (or first law) is obtained by multiplying Eq. (\ref{EqLKQ}) by $\dot x_t$ and integrating over the time interval $[0,t]$. This yields\footnote{Throughout this paper, products of stochastic variables and stochastic integrals are defined within the Stratonovich interpretation and denoted by $\circ$.}
\begin{align}
\Delta E=w[\{x_s\}]-q[\{x_s\}]
\end{align}
where 
\begin{align}
\Delta E=\int_0^t ds\: [m\ddot x_s+kx_s] \circ \dot x_s=\frac{m}{2}[v_t^2-v_0^2]+\frac{k}{2}[x_t^2-x_0^2]
\end{align}
is the change in the internal energy of the (Brownian) system, 
\begin{align}
w[\{x_s\}]&=\int_0^t ds\: F_{\textnormal{fb}}(\dot x_s)\circ \dot x_s\nonumber\\
&= -\gamma' \int_0^t ds\: \dot x_s^2
\end{align}
is the work done by the feedback force on the system, and 
\begin{align}
\label{EqDeltasm}
q[\{x_s\}]&= \int_0^t ds\:[\gamma \dot x_s-\sqrt{2\gamma T}\xi_s]\circ \dot x_s \nonumber\\
&=-\int_0^t ds\: [m \ddot x_s+\gamma' \dot x_s +kx_s] \circ \dot x_s  
\end{align}
is the heat dissipated into the thermal environment\footnote{By convention, we assign a positive sign to $q[\{x_s\}]$ if the energy is dissipated into the bath.} which can also be identified with an entropy increase in the medium $\Delta s_m[\{x_s\}]\equiv q[\{x_s\}]/T$.
The crucial point is that this quantity can also be written in the form\cite{MR2012}
\begin{align}
\label{Eqratio0}
\Delta s_m[\{x_s\}]=\ln \frac{{\cal P}_+[\{x_s\}\vert x_0,v_0]}{{\cal P}_-[\{\hat x_s\}\vert \hat x_0,\hat v_0]}-\frac{\gamma'}{m} t
\end{align}
where  
\begin{align}
\label{Eqpath0}
{\cal P}_{+}[\{x_s\}\vert x_0,v_0]&\propto \exp\left[\frac{\gamma+\gamma'}{2m} t-\frac{1}{4\gamma T}\int_0^t ds \:(m\ddot x_s+(\gamma +\gamma')\dot x_s+kx_s)^2\right]
\end{align}
is the conditional weight of the path $\{x_s\}_{s\in [0,t]}$ given  the initial point $(x_0, v_0)$ and ${\cal P}_-[\{\hat x_s\}\vert \hat x_0,\hat v_0]$ is the conditional weight of the time-reversed path $\{\hat x_s\}_{s\in [0,t]}$ (defined by $\hat x_s\equiv x_{t-s},\dot {\hat x}_s\equiv -\dot x_{t-s}$) generated by the ``conjugate" Langevin equation in which $\gamma'$ is replaced by $-\gamma'$. The additional term 
\begin{align}
\Delta s_{\textnormal{pu}}=-\frac{\gamma'}{m} t
\end{align}
is interpreted as an ``entropy pumping"   arising from the contraction of momentum phase space due to the feedback force\footnote{To trace back the origin of the term $(\gamma+\gamma')t/(2m)$ in the path probability ${\cal P}_{+}[\{x_s\}\vert x_0,v_0]$, see  e.g. \cite{HR1981,IP2006,SF2012}. The normalization factor  cancels out when taking the ratio of the two probabilities.}. This is a unique feature of a velocity-dependent feedback control\cite{KQ2004,KQ2007}. As stressed in \cite{MR2012}, the fact that $\gamma' $ must be treated as an odd variable under time reversal in order to relate $\Delta s_m[\{x_s\}]$ to the microscopic irreversibility of trajectories is not harmless: it implies that there is no steady state with the conjugate dynamics when $\gamma'>\gamma $, which is the common situation encountered in cold damping setups. 

Introducing the stochastic entropy of the Brownian system\cite{C1999,Q2001,S2005},
\begin{align}
s_{\textnormal{sys}}(t)=-\ln p_t(x_t,v_t)
\end{align}
where $p_t(x_t,v_t)$ is the solution of the Fokker-Planck equation evaluated along the trajectory, the entropy production for each realization of the stochastic process in the time interval $[0,t]$ is then defined as
\begin{align}
\label{Eqsigmaxs}
\sigma [\{x_s\}]\equiv \ln \frac{{\cal P}_+[\{x_s\}]}{{\cal P}_-[\{\hat x_s\}]}=\Delta s_{\textnormal{sys}}+\Delta s_m[\{x_s\}]-\Delta s_{\textnormal{pu}}
\end{align}
where
\begin{align}
\label{EqDeltas}
\Delta s_{\textnormal{sys}}=\ln \frac{p_0(x_0,v_0)}{p_t(x_t,v_t)} \ .
\end{align}
 As a direct consequence of Eqs. (\ref{Eqratio0}) and (\ref{EqDeltas}), $\sigma [\{x_s\}]$ obeys the integral fluctuation theorem (IFT) 
\begin{align}
\label{EqIFT0}
 <e^{-\sigma[\{x_s\}]}> =1  
\end{align}
where $< ...> $ denotes a  functional average over all  paths $\{x_s\}_{s\in[0,t]}$ generated by Eq. (\ref{EqLKQ}) that start from $(x_0,v_0)$, as well as averages over initial and final positions and velocities. One can also derive a detailed FT which has a nontrivial interpretation in the NESS because of the change of sign of $\gamma'$, as discussed in \cite{MR2012}. 
From Jensen inequality, it follows from Eq. (\ref{EqIFT0}) that the average of $\sigma [\{x_s\}]$ is a non-negative quantity and thus
\begin{align}
\label{Eq2ndL}
\ <\Delta s_{\textnormal{sys}}+\Delta s_m[\{x_s\}]> \: \ge   \Delta s_{\textnormal{pu}}\ .
\end{align}
In other words, the  total variation of entropy in the system plus the bath may be negative on average (which may look as a violation of the second law of thermodynamics if the role of the external agent is ignored) but this quantity is always bounded from below by $ -(\gamma/m)t$. In this sense, $- \Delta s_{\textnormal{pu}}$  plays a role similar to mutual information in the generalization of the second law to systems under feedback control\cite{SU2012}.  

Upon averaging,  $\Sigma \equiv < \sigma [\{x_s\}]>$, $Q\equiv <q[\{x_s\}]>$, and $\Delta S_{m}\equiv Q/T\equiv < \Delta s_{m}[\{x_s\}]>$  become the non-equilibrium thermodynamic quantities defined at the ensemble level, and the corresponding averaged rates are given by\cite{KQ2004}
\begin{align}
\label{EqPERP}
\dot \Sigma(t)=\frac{1}{\gamma T} \int dx dv \frac{J_t^2(x,v)}{ p_t(x,v)}  \ ,
\end{align}
\begin{subequations}
\label{EqSm}
\begin{align}
\label{EqSma}
\dot S_{m}(t)&\equiv \frac{\dot Q}{T}=-\frac{1}{T} \int dx dv \: v  J_t(x,v)\\
\label{EqSmb}
&= \frac{\gamma}{T} [<v^2>_t-\frac{T}{m}] 
\end{align}
\end{subequations}
where $<v^2>_t\equiv \int dx dv v^2 p_t(x,v)$. The  entropy balance equation (or generalized second law) then takes the form
\begin{align}
\label{Eqbal0}
\dot \Sigma(t)=\dot S_{\textnormal{sys}}(t)+\dot S_m(t)-\dot S_{\textnormal{pu}} \ge 0
\end{align}
with 
\begin{align}
\dot S_{\textnormal{pu}}=-\frac{\gamma'}{m} \ .
\end{align}
This  equation can also be derived directly by taking the time derivative of the system Gibbs-Shannon entropy 
\begin{align}
\label{EqSsys}
S_{\textnormal{sys}}(t)\equiv <s_{\textnormal{sys}}(t)>_t=-\int dx dv \: p_t(x,v) \ln  p_t(x,v) 
\end{align}
and inserting the Fokker-Planck equation (\ref{EqFPKQ}). 

Note that $J_t(x,v)$, as defined by Eq. (\ref{EqJ0}), is actually the  {\it irreversible} component of the total probability current\cite{R1989} since $\gamma'$ is odd  under time reversal. Therefore, Eqs.  (\ref{EqPERP})  and (\ref{EqSma}) are in agreement with the general definitions of the non-negative irreversible entropy production and of the heat flow in a stochastic system with odd and even variables (see e.g. \cite{SF2012}).  

The rates  $\dot \Sigma$ and $\dot S_{m}$ have simple expressions in the NESS where the solution of the FP equation has the form of an equilibrium Gibbs distribution 
\begin{align}
\label{EqPst}
p_{\textnormal{st}}(x,v)=  \frac{\sqrt{km}}{2\pi T_{\textnormal{eff}}}e^{-\frac{kx^2+mv^2}{2T_{\textnormal{eff}}}}
\end{align}
with an {\it effective} temperature lower than $T$
\begin{align}
T_{\textnormal{eff}}\equiv m<v^2>_{\textnormal{st}}=\frac{\gamma}{\gamma +\gamma'}T \ .
\end{align}
Then
\begin{subequations}
\begin{align}
\label{EqdotSigmaa}
\dot \Sigma&=\dot Q(\frac{1}{T}-\frac{1}{T_{\textnormal{eff}}})\\
\label{EqdotSigmab}
&=\frac{\gamma'^2}{m(\gamma+\gamma')} \ ,
\end{align}
\end{subequations}
and
\begin{subequations}
\begin{align}
\label{EqdotSma}
\dot S_m&\equiv \frac{\dot Q}{T}=\frac{\gamma}{m} \frac{T_{\textnormal{eff}}-T}{T}\\
\label{EqdotSmb}
&=-\frac{\gamma \gamma'}{m(\gamma+\gamma')} \ .
\end{align}
\end{subequations}
 Hence, heat  flows from the reservoir to the system on average, and Eq.(\ref{EqdotSigmaa}) merely describes the entropy flux between two objects at temperature $T$ and $T_{\textnormal{eff}}$. 

\section{Entropy production and entropy pumping with measurement errors}

We now build on the analysis of the preceding section to study the EP in models V and P described by Eqs. (\ref{EqmodelV}) and  (\ref{EqmodelP}), respectively. 
Thanks to the linearity of the Langevin equations, the probability distributions and the power spectral densities can be explicitly calculated in the NESS and their  expressions are given in Appendix A.

\subsection{Model V in the limit of a Markovian feedback}

We first study model V in the limit  $\tau=0$, that is when the feedback force is directly proportional to the output signal of the detector 
\begin{align}
F_{\textnormal{fb}}(y)&=-\gamma' y(t)\nonumber\\
&=-\gamma' [\dot x(t)+\sqrt{S_{v_n}} \eta (t)]
\end{align}
and the Langevin equation reads 
\begin{align}
\label{EqLKQ2}
m\ddot x +\gamma \dot x+\gamma' [\dot x+ \sqrt{S_{v_n}} \eta (t)]+ kx=\sqrt{2\gamma T}\: \xi(t) \ .
\end{align}

The first question that arises is whether the previous analysis at the level of an individual stochastic trajectory $\{x_s\}_{s\in [0,t]}$ can be  generalized.  The key feature is that the  probability of the trajectory only contains the {\it total} noise acting on the system. Indeed, since the sum of two independent Gaussian noises is also a Gaussian noise, the  trajectories generated by Eq. (\ref{EqLKQ2}) are also generated by the  Langevin equation  
\begin{align}
\label{EqLKQ3}
m\ddot x +(\gamma +\gamma')\dot x+kx=\sqrt{2(\gamma T+\gamma' T')}\: \rho(t) \ ,
\end{align}
where $T'=\gamma' S_{v_n}/2$  has the dimension of temperature and  $\rho(t)$ is a zero-mean, delta-correlated noise $<\rho(t)\rho(t')>=\delta (t-t')$. As a result, the conditional path probability ${\cal P}_+[\{x_s\}\vert x_0,v_0]$ is given by Eq. (\ref{Eqpath0}) with $T$ replaced by  $(\gamma T+\gamma'T')/\gamma$. It follows that
\begin{align}
\label{Eqfaux}
\ln \frac{{\cal P}_+[\{x_s\}]}{{\cal P}_-[\{\hat x_s\}]}=\Delta s_{\textnormal{sys}}-\frac{\gamma}{\gamma T+\gamma' T'}\int_0^t ds\: \left[m \ddot x_s+\gamma' \dot x_s +kx_s\right] \circ \dot x_s + \frac{\gamma'}{m} t 
\end{align}
(noting that the product $\gamma' T'=\gamma'^2 S_{v_n}/2$ is no affected by the change $\gamma'$ to $-\gamma'$). The  problem is that this logratio cannot be considered as a sensible definition of the entropy production $\sigma[\{x_s\}]$ along the trajectory. In the first place, the second term in the right-hand side does not identify with $\Delta s_m[\{{\bf X}_s\}]\equiv q[\{{\bf X}_s\}]/T$, the entropy change in the medium, where the exchanged heat is defined in the usual way  as\footnote{In this paper, following \cite{Sbook2010,Seki1997}, we define the heat as the work done by the ``reaction" force $\gamma \dot x_t-\sqrt{2\gamma T}\xi_t$  on the surrounding fluid due to the motion of the Brownian entity. This definition does not change in presence of a measurement noise.} 
 \begin{align}
\label{EqLheat}
q[\{{\bf X}_s\}]&=\int_0^t ds\:[\gamma \dot x_s-\sqrt{2\gamma T}\xi_s]\circ \dot x_s \nonumber\\
&=-\int_0^t ds\: \left[m \ddot x_s+\gamma' y_s +kx_s\right] \circ \dot x_s  \ .
\end{align}
where $y_s=\dot x_s+\sqrt{S_{v_n}}\eta_s$. As indicated, $q$ and $\Delta s_m$ are now functionals of $\{{\bf X}\}_{s\in[0,t]}\equiv (\{x_s\},\{y_s\})_{s\in[0,t]}$ (or functionals of the two  noises $\{\xi_s\}$ and $\{\eta_s\}$). In the second place, the average of Eq. (\ref{Eqfaux}) in the NESS does not depend on the measurement noise. Indeed, since  the stationary probability distribution is again given by Eq. (\ref{EqPst}) with an effective temperature
\begin{align}
\label{EqTeff1}
T_{\textnormal{eff}}\equiv m<v^2>_{\textnormal{st}}=\frac{\gamma}{\gamma+\gamma'} T+\frac{\gamma'} {\gamma+\gamma'}T' \ ,
\end{align}
Eq. (\ref{Eqfaux}) yields
\begin{align}
\label{Eqratio0V}
\frac{1}{t}<\ln \frac{{\cal P}_+[\{x_s\}}{{\cal P}_-[\{\hat x_s\}]}>_{\textnormal{st}}&=-\frac{\gamma\gamma'}{\gamma T+\gamma' T'}< v^2>_{\textnormal{st}} + \frac{\gamma'}{m}\nonumber\\
&=\frac{\gamma'^2}{m(\gamma +\gamma')}\ ,
\end{align}
which is the same as  Eq. (\ref{EqdotSigmab}) for an error-free measurement\footnote{Since the probability of a trajectory is Gaussian in the NESS, it is of course crucial to change the sign of $\gamma'$ when time is reversed. If not, the logratio would just predict a vanishing entropy production rate. This special property of linear Langevin equations will be encountered again in the next section and is discussed in detail in \cite{CPV2012} in the overdamped case.}. Therefore, defining the entropy production from the microscopic irreversibility of the trajectories $\{x_s\}_{s\in[0,t]}$ is not pertinent in the present context (alternatively, one could consider the probability of  $\{{\bf X}\}_{s\in[0,t]}$, but this amounts to changing the level of description of the system since $y_s$ is no more a ``hidden" variable, as will be discussed below and in more detail in the next section).

To bypass this difficulty and still define the EP and the entropy pumping for $T'>0$, there is no other choice than to work at the ensemble level from the outset as was done originally in \cite{KQ2004}. 
To this end,  the different terms in the Fokker-Planck equation must be rearranged appropriately. From Eq. (\ref{EqLKQ3}), this equation reads
\begin{align}
\label{EqFPKQn1}
\partial_t  p_t(x,v)=-\partial_x [vp_t(x,v)] +\frac{1}{m}\partial_v\left[[kx+(\gamma+\gamma')v]p_t(x,v) +\frac{\gamma T+\gamma' T'}{m} \partial_v p_t(x,v)\right] \ ,
\end{align}
which is conveniently rewritten as 
\begin{align}
\label{EqFPKQn2}
\partial_t  p_t(x,v)=-\partial_x [vp_t(x,v)] -\frac{1}{m}\partial_v\left[[-kx+\tilde F_{\textnormal{fb}}(v,t)] p_t(x,v)+ J_t(x,v)\right]
\end{align}
where $J_t(x,v)$ is defined by Eq. (\ref{EqJ0})  and 
\begin{align}
\label{EqFeff}
\tilde F_{\textnormal{fb}}(v,t)=-\gamma' [ v+\frac{T'}{m} \partial_v \ln p_t(x,v)] 
\end{align}
plays the role of an effective (or {\it apparent}) feedback force. In this form, the FP equation is quite similar to Eq. (\ref{EqFPKQ}) for $T'=0$, that is for an error-free measurement. In particular, the probability current  $J_t(x,v)$ keeps the same  definition, which is justified by the fact that Eq. (\ref{EqSma}) still gives the correct result for the average heat $Q=T\Delta S_m=<q[\{x_s\},\{y_s\}]>$, where  the average is taken over all possible realizations $\{\xi_s\}$ and $\{\eta_s\}$ of the noises in the time interval $[0,t]$\footnote{This is because the Gaussian noises $\xi(t)$ and $\eta(t)$ are independent. The average of the product $\xi_s\circ \dot x_s$ in the first line of Eq. (\ref{EqLheat}) is then  equal to $\sqrt{2\gamma T}/(2m)$, as can be shown by using Novikov's theorem\cite{N1964} for instance. Therefore Eq. (\ref{EqSmb}) remains true and this result is also obtained from Eq. (\ref{EqSma}).}.

Accordingly, the entropy balance equation obtained by taking the time derivative of the Gibbs-Shannon entropy is formally the same as for $T'=0$,
\begin{align}
\label{Eq2ndL3}
\dot {\tilde \Sigma}(t)=\dot S_{\textnormal{sys}}(t)+\dot S_m(t)-\dot S_{\textnormal{pu}}(t) \ge 0\ ,
\end{align}
%where $\dot S_m(t)$ and $\dot {\tilde \Sigma}(t)$ are given by Eqs. (\ref{EqSm}) and (\ref{EqPERP}), respectively, and
where  ${\tilde \Sigma}(t)$, the non-negative  {\it apparent} entropy production rate\footnote{The use of the term ``apparent " borrowed from \cite{MLBBS2012} will be justified below in section 4B}, is given by Eq. (\ref{EqPERP}), $\dot S_m(t)$  is given by Eq. (\ref{EqSm}), and
\begin{align}
\label{EqSpu}
\dot S_{\textnormal{pu}}(t)&=\frac{1}{m}\int dx dv \: p_t(x,v)\frac{\partial \tilde F_{\textnormal{fb}}(v,t)}{\partial v}\nonumber\\
&= -\frac{\gamma'}{m} [1+\frac{T'}{m}\int dx dv \: p_t(x,v)\frac{\partial^2\ln p_t(x,v)}{\partial v^2}] \ .
\end{align}
Eq. (\ref{EqSpu}) generalizes the entropy pumping in  presence of measurement errors and is the main result of this section (as it must be, the results of \cite{KQ2004} recalled in  section 3 are recovered for $T'=0$, with $\dot {\tilde \Sigma} \rightarrow \dot \Sigma$).  In general, the entropy pumping is time-dependent and the physical meaning of the second term in the right-hand side of Eq. (\ref{EqSpu}) is not transparent. Things become more intelligible in the NESS as the probability distribution $p_{\textnormal{st}}(x,v)$ has again the form of a Gibbs-Boltzmann distribution with $T_{\textnormal{eff}}$ given by (\ref{EqTeff1}). The apparent feedback force (\ref{EqFeff}) is then proportional to the instantaneous velocity
\begin{align}
\label{EqFeffst}
\tilde F_{\textnormal{fb,st}}(v)=-\tilde \gamma' v
\end{align}
with an apparent damping coefficient 
\begin{subequations}
\begin{align}
\label{Eqgammaeffa}
\tilde \gamma'&=\gamma' \frac{T_{\textnormal{eff}}-T'}{T_{\textnormal{eff}}}=\gamma \frac{T-T_{\textnormal{eff}}}{T_{\textnormal{eff}}}\\
\label{Eqgammaeffb}
&=\gamma \gamma' \frac{T-T'}{\gamma T+\gamma' T'} \ .
\end{align}
\end{subequations}
Eqs. (\ref{EqPERP}), (\ref{EqSma}), and (\ref{EqSpu}) then yield

\begin{subequations}
\label{EqdSigma}
\begin{align}
\label{EqdSigmaa}
\dot {\tilde \Sigma}&=\dot Q(\frac{1}{T}-\frac{1}{T_{\textnormal{eff}}})\\
\label{EqdSigmab}
&=\frac{\gamma \gamma'^2}{m(\gamma +\gamma')} \frac{(T-T')^2}{(\gamma T+\gamma' T')T}  \ ,
\end{align}
\end{subequations}
\begin{subequations}
\label{EqdSm}
\begin{align}
\label{EqdSma}
\dot S_m&\equiv\frac{\dot Q}{T}=\frac{\gamma}{m} (T_{\textnormal{eff}}-T)\\
\label{EqdSmb}
&=-\frac{\gamma \gamma'}{m(\gamma +\gamma')} \frac{T-T'}{T} \ ,
\end{align}
\end{subequations}
and
\begin{subequations}
\label{Eqdspu}
\begin{align}
\label{Eqdspua}
\dot S_{\textnormal{pu}}&=-\frac{\tilde \gamma'}{m}\\
\label{Eqdspub}
& =-\frac{\gamma \gamma' }{m}\frac{T-T'}{\gamma T+\gamma' T'}\ .
\end{align}
\end{subequations}
Hence, the picture of heat exchange between two objects at temperature $T$ and $T_{\textnormal{eff}}$ is still pertinent in the NESS, with the measurement noise only increasing the value of $T_{\textnormal{eff}}$ and thus reducing the efficiency of the cooling\cite{PZ2012}.  Eq. (\ref{Eqgammaeffa}) shows that this can also be interpreted as a weakening of the apparent feedback force, which in turn decreases the average heat flow coming from the reservoir and the apparent EP. We also remark that $-\dot S_{\textnormal{pu}}$ decreases as $T'$  increases\footnote{Eq. (\ref{Eqdspub}) can be written as $-\dot S_{\textnormal{pu}}=\frac{\gamma'}{m} -\frac{\gamma'}{m} \frac{\gamma + \gamma'}{\gamma T + \gamma' T'}T'$.}. Since $\dot S_m - \dot S_{\textnormal{pu}}\ge 0$ from Eq. (\ref{Eq2ndL3}) (as $\dot S_{\textnormal{sys}}=0$ in the NESS by definition), the standard formulation of the second law is less and less ``violated" as the measurement error increases, which again shows that entropy pumping plays a role similar to mutual information in the generalization of the second law\footnote{Note that Eqs. (52)-(54) are  not specific to a harmonic model and are more universally valid (e.g., if one adds an additional quartic term in the potential) as long as the feedback force is proportional to the output signal of the detector, as given by Eq. (38). The stationary probability distribution is then still a Gibbs-Boltzmann distribution and the effective temperature is defined by Eq. (43). In particular, the apparent feedback force keeps the suggestive form of Eq. (50).}

\begin{figure}[hbt]
\begin{center}
\includegraphics[width=12cm]{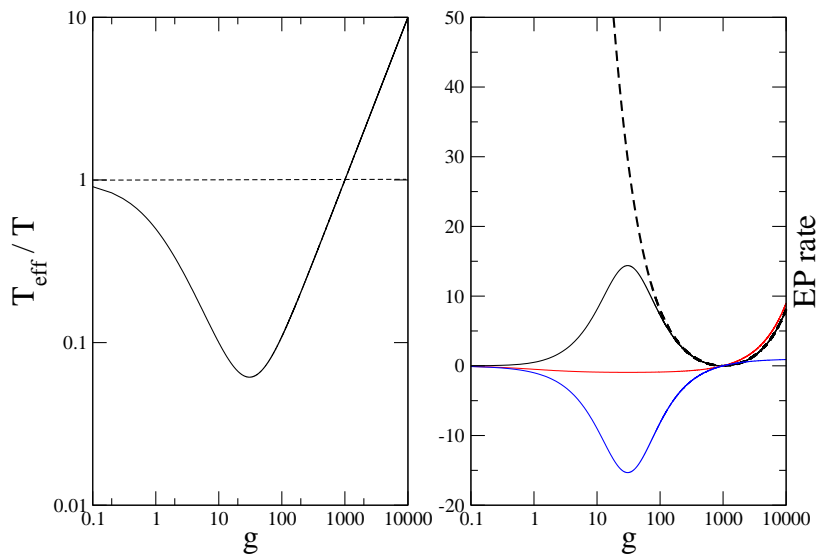}
\caption{\label{Fig1} (a)  Resonator temperature $T_{\textnormal{eff}}/T$ as a function of  the feedback gain $g=\gamma'/\gamma$ for a signal-to-noise ratio SNR$=2T/(\gamma S_{v_n})=1000$. (b) Different contributions to the entropy production rate according to Eqs. (\ref{EqdSigmab}), (\ref{EqdSmb}), and (\ref{Eqdspub}) (in units $k_B\tau_0^{-1}$): $\dot {\tilde \Sigma}$ (black solid line), $\dot S_m$ (red solid line), $\dot S_{\textnormal{pu}}$ (blue solid line). The dashed line is the total entropy production rate given by Eq. (\ref{Eqsigmatotb}) when the Langevin equation (\ref{EqLKQ2}) is viewed as describing a Brownian particle coupled to two heat reservoirs.}
\end{center}
\end{figure}

As shown in Fig. 1, the apparent EP rate $\dot {\tilde \Sigma}$ has an interesting behavior as a function of the feedback gain $g=\gamma'/\gamma$, in relation with the behavior of the resonator temperature $T_{\textnormal{eff}}$. As is well known in the theory of feedback cooling\cite{PZ2012,PDMR2007} and is indeed observed experimentally (see for instance \cite{VBMF2010} for a setup related to model V), the temperature $T_{\textnormal{eff}}$  reaches a minimum at a certain value $g_{min}$ of the gain  for a given  signal-to-noise ratio (SNR)\footnote{Eq. (\ref{EqTeff1}) yields $g_{min}=\sqrt{1+{\textnormal SNR}}-1$ and thus $T_{\textnormal{eff,min}}=2T(\sqrt{1+{\textnormal SNR}}-1)$, where SNR$\equiv S_{vv}^{g=0}(\omega_0)/{S_{v_n}}$ is the ratio of the original thermal noise peak (i.e., without feedback) to the detector noise floor. From Eq. (\ref{EqPSDVa}) with $\tau=0$, SNR$=2T/(\gamma S_{v_n})$. Hence $T'/T=g/$SNR.}. Above $g_{min}$, too much detector noise is fed back to the resonator and $T_{\textnormal{eff}}$ starts to increase. In particular,  $T_{\textnormal{eff}}=T$ for $g=2T/(\gamma S_{v_n})=$SNR (which corresponds to $T'=T$): the resonator is then at equilibrium with its environment, as if there were no feedback.  For even larger values of $g$, the resonator is  heated. Interestingly, Eqs. (\ref{EqdSigma}) tell us that the apparent EP rate is maximal when $T_{\textnormal{eff}}$ is minimal and, as can be seen in Fig. 1, the essential contribution to the EP around $g_{min}$ comes from entropy pumping ($\vert \dot S_{\textnormal{pu}}/\dot S_m/\vert_{g=g_{min}}\approx \sqrt{{\textnormal SNR}}/2$). Loosely speaking, one may say that the optimal cooling is achieved when the controller extracts the maximal information about the state of the system via the  measurement. This can also be put in relation with the behavior of the spectral density $S_{v'v'}(\omega)$ of the output signal $v'=v+v_n$, as noticed in \cite{VBMF2010}.  Finally, we stress that the entropy pumping vanishes for $T=T'$ like the heat flow from the reservoir, so that there is no EP on average, as the heating due to the detector noise exactly compensates the cooling due to the extra friction.

At this stage, it is interesting to compare the above results to those obtained when Eq. (\ref{EqLKQ2}) is regarded as a Langevin equation describing a Brownian particle coupled to two thermal environments at  temperatures $T$ and $T'$. This is a model  (with or without the harmonic trap) which is often discussed in the literature\cite{PE1996,VdbKM2004,DB2005,V2006,VdBE2010,FI2011,S2012} (see also \cite{Sbook2010} and the recent experimental work described in \cite{CINT2013}) as it is probably the simplest example of heat conduction.
In this interpretation, the heat flowing from each thermostat to the particle drives the system out of equilibrium, and to correctly distinguish the two heat flows the Fokker-Planck equation (\ref{EqFPKQn1}) is written as 
\begin{align}
\label{EqFPKQn}
\partial_t  p_t(x,v)=-\partial_x [vp_t(x,v)] -\frac{1}{m}\partial_v\left[ -kx p_t(x,v)+ J_t(x,v)+J'_t(x,v)\right]
\end{align}
where
\begin{align}
\label{Eqjjp}
J'_t(x,v)=-\gamma' [ v+\frac{T'}{m} \partial_v \ln p_t(x,v)] p_t(x,v)  \ .
\end{align}
This leads to the entropy balance equation (see e.g. \cite{VdBE2010})
\begin{align}
\label{Eq2ndL1}
\dot \Sigma(t) =\dot S_{\textnormal{sys}}(t)+ \dot S_{m}(t)+\dot S_{m'}(t) \ ,
\end{align}
where $\dot S_{m'}(t)$, the entropy flow to the second reservoir, is defined like $\dot S_{m}(t)$  (with $T$ and $J_t(x,v)$ replaced by $T'$ and $J'_t(x,v)$, respectively), and
\begin{align}
\label{EqPERPtot}
\dot \Sigma(t)= \frac{1}{\gamma T}\int dx dv \:\frac{J_t^2(x,v)}{p_t(x,v)}+\frac{1}{\gamma' T'}\int dx dv\:\frac{J_t'^2(x,v)}{p_t(x,v)} \ .
\end{align}
In the steady state, this yields
\begin{subequations}
\label{Eqsigmatot}
\begin{align}
\label{Eqsigmatota}
\dot \Sigma&=\dot Q(\frac{1}{T}-\frac{1}{T'})\\
\label{Eqsigmatotb}
&=\frac{\gamma \gamma'}{m(\gamma + \gamma')}\frac{(T-T')^2}{TT'}
\end{align}
\end{subequations}
where
\begin{align}
\label{EqSmSm'}
\dot Q&=T\dot S_m=\frac{\gamma}{m} (T_{\textnormal{eff}}-T)\nonumber\\
&=-T'\dot S_{m'}=-\frac{\gamma'}{m} (T_{\textnormal{eff}}-T') \ .
\end{align}
Eq. (\ref{Eqsigmatotb}) is the usual thermodynamic expression for the average EP rate associated with a steady heat flux between two reservoirs mediated by a device with thermal conductivity $\gamma \gamma'/[m(\gamma + \gamma')]$ (see e.g \cite{PE1996,VdbKM2004,VdBE2010}). This expression is quite different from the average apparent EP rate given Eq. (\ref{EqdSigmab}) (whereas $\dot Q$ does not change). Indeed, Eq. (\ref{EqdSigmaa})  has not the same physical content as Eq. (\ref{Eqsigmatota}) since only the two temperatures $T$ and $T_{\textnormal{eff}}$ come into play in the former case  and $T'$ is only introduced for convenience (changing $T'$ changes the value of $T_{\textnormal{eff}}$). In other words, at the level of description corresponding to Eq. (\ref{EqdSigma}), the entropy exchange with the external agent is not associated with another heat flow  but only manifests itself in the form of entropy pumping. This makes a big difference since the average EP rate given by Eq. (\ref{EqdSigmab}) remains finite when the measurement becomes error-free ($T'\rightarrow 0$) whereas $\dot S_{m'}=\dot Q/T'$ and thus $\dot \Sigma$ given by Eq. (\ref{Eqsigmatotb}) diverge\footnote{More generally, the EP rate given by Eq. (\ref{EqPERP}), which leads to Eq. (\ref{EqdSigmab}) in the NESS, is always smaller than the EP given by Eq. (\ref{EqPERPtot}) since the contribution involving $J_t'^2(x,v)/p_t(x,v)$ is not present. In the NESS, the difference between the two EP rates become asymptotically zero in the large $g$ limit, as can be seen in Fig. 1.} (see \cite{SSBE2012} for a closely related discussion in the case of a true Maxwell demon). Moreover,  as a function of $g$, $\dot \Sigma$  does not display any extremum at $g=g_{min}$, which shows that this quantity does not reflect the interesting physics of the problem. More fundamentally, there is no rationale for considering $T'$ as the genuine temperature of another reservoir in the context of cold damping.
It it worth noting, however,  that Eq. (\ref{Eqsigmatotb}) is also the expression of the average EP resulting from the analysis  performed in \cite{K2012}, as can be readily checked. In this work,  the  Langevin Eq. (\ref{EqLKQ2}) (with $k=0$) is obtained by taking a suitable continuous-time limit of a discrete series of independent measurements of the velocity (as originally considered in \cite{IS2011})\footnote{\label{Kundu} Specifically, the conditional probability for obtaining the measurement outcome $y(t)$ from the velocity $v(t)$ is given by a Gaussian distribution with  variance $\Delta^2 \delta t$ where $\Delta$ quantifies the error and $\delta t$ is the infinitesimal time step. The limit $\Delta \rightarrow \infty$ and $\delta t \rightarrow 0$ is then taken such that $\Delta^2 \delta t \rightarrow 2\alpha_0$ finite. Hence $2\alpha_0$  identifies with the noise spectral density $S_{v_n}$ in Eq. (\ref{EqLKQ2}).}. A quantity that identifies with $\Delta S_{m'}$ (the change in entropy of the second ``heat" reservoir) is then interpreted as the ``entropy production due to the measurement process" and contributes to the total EP. In the light of the above discussion, it is clear that this corresponds to another level of description of the system. We shall come back to this issue in the next section and in Appendix C.

Now that we have defined the average (apparent) EP rate associated with the Fokker-Planck equation (\ref{EqFPKQn2}),  let us again consider individual trajectories and see whether we can define a corresponding (apparent) EP functional. The main problem is to relate the entropy pumping to a momentum phase space contraction like in the case $T'=0$. To simplify the discussion, and also because this is the normal regime of a cold damping setup, we only consider stationary trajectories.  Then, Eq. (\ref{EqFeffst}) immediately suggests to replace the original Langevin equation (\ref{EqLKQ2}) by the {\it effective}  equation 
\begin{align}
m\ddot x +(\gamma +\tilde \gamma')\dot x+kx=\sqrt{2\gamma T}\: \xi(t) \
\end{align}
which can be rewritten, using Eq. (\ref{Eqgammaeffa}), as
\begin{align}
\label{EqLeff}
m\ddot x +\gamma\left(\frac{T}{T_{\textnormal{eff}}}\right)\dot x+kx=\sqrt{2\gamma T}\: \xi(t) \ .
\end{align}
By construction, this equation leads to a NESS with the same probability distribution $p_{\textnormal{st}}(x,v)$ as Eq. (\ref{EqLKQ2}). On the other hand, the individual stochastic trajectories are different. Since only the Langevin thermal noise $\xi(t)$ appears in Eq. (\ref{EqLeff}), we are led back to the problem treated in \cite{KQ2004,KQ2007,MR2012} and recalled in section 3. Hence,  the {\it apparent} entropy change  $\tilde \Delta s_m[\{x_s\}]\equiv \tilde q[\{x_s\}]/T$, where $\tilde q[\{x_s\}]$ is the {\it apparent}  heat dissipated in the environment  defined by  
\begin{align}
\tilde q[\{x_s\}]&=-\int_0^t ds\: [m \ddot x_s+\tilde \gamma'\dot x_s +kx_s] \circ \dot x_s  \ ,
\end{align}
also satisfies Eq. (\ref{Eqratio0}) with $\gamma'$ replaced by $\tilde \gamma'$ everywhere. It follows that the {\it apparent} path-dependent EP 
\begin{align}
\label{Eqsigmatilde}
\tilde \sigma [\{x_s\}]\equiv \ln \frac{{\cal \tilde P}_+[\{x_s\}]}{{\cal \tilde P}_-[\{\hat x_s\}]}=\Delta s_{\textnormal{sys}}+ \Delta \tilde  s_m[\{x_s\}]+\frac{\tilde \gamma'}{m}t \ ,
\end{align}
where  ${\cal \tilde P}_+[\{x_s\}]$ (resp. ${\cal \tilde P}_-[\{x_s\}]$) is the probability of a path $\{x_s\}_{s\in [0,t]}$ generated by Eq. (\ref{EqLeff}) in the NESS (resp. by the conjugate equation with $\tilde \gamma'$ replaced by $-\tilde \gamma'$), satisfies the IFT $ <e^{- \tilde \sigma[\{x_s\}]}>_{eff,st} =1$. A proper detailed fluctuation theorem can be also obtained following \cite{MR2012}. By construction,  $(1/t)<\tilde \sigma[\{x_s\}]>_{eff,st}$ and $(1/t)<\tilde q[\{x_s\}]>_{eff,st}$ identify with the  average apparent EP rate and the average heat flow $\dot Q$ given by Eqs.  (\ref{EqdSigmab}) and (\ref{EqdSmb}), respectively.

%To estimate the error that we make in replacing Eq. (\ref{EqLKQ2}) by Eq. (\ref{EqLeff}), we compare in Appendix .... the corresponding power strectra $S_{vv}(\omega)$ and the probability distributions of $q[\{x_s\}]$ and  $\tilde q[\{x_s\}]$ which can be calculated exactly.  Of course, in practice, the error depends on the numerical values of the system parameters, as will be discussed in Section....

Let us finally remark that the effective Langevin equation can also be derived from the original Langevin equation (\ref{EqLKQ2}) by  replacing the measurement noise $\eta(t) $ by its projection onto the space spanned by the stochastic variables $x$ and $v$, which is defined as
\begin{align}
\tilde \eta_{\textnormal{st}}(x,v)=\frac{<\eta \circ x>}{<x^2>_{\textnormal{st}}}x+\frac{<\eta \circ v>}{<v^2>_{\textnormal{st}}}v \ .
\end{align}
Since  $<\eta \circ x>=0$ and $<\eta \circ v>=-\sqrt{2\gamma' T'}/(2m)$, one has $\tilde \eta_{\textnormal{st}}=\sqrt{2\gamma' T'}/(2T_{\textnormal{eff}})\: v$ and  Eq. (\ref{EqLeff}) is recovered by inserting this result into Eq. (\ref{EqLKQ2}). This procedure will be useful to derive similar effective Langevin equations in what follows.

\subsection{Model V for $\tau >0$ }

We now generalize the above analysis to the  case $\tau >0$. Whereas Eqs. (\ref{EqmodelV})  describe the coupled dynamics of the two processes $x(t)$ and $y(t)$,  we are  interested in the apparent entropy production associated with $x(t)$ only. Like in the overdamped case considered in \cite{ CPV2012}, the coupling makes the effective dynamics of $x(t)$ no longer Markovian, as we have  already noticed. However, since  the Langevin equations are linear, the path probability ${\cal P}_+[\{x_s\}]$ is  Gaussian in the NESS, and the ratio ${\cal P}_+[\{x_s\}]/{\cal P}_-[\{\hat x_s\}]$ can be easily computed by going to Fourier space, as shown in Appendix B. It turns out that the average of this quantity is independent on the measurement error, like for $\tau=0$. Therefore, again, we cannot define an apparent EP functional from the microscopic irreversibility of the trajectories $\{x_s\}$ and we need to first derive the entropy balance equation at the ensemble level.

We thus consider  the time evolution of the Shannon entropy $S_{\textnormal{sys}}(t)=-\int dx dv \: p_t(x,v)\ln p_t(x,v)$ where $p_t(x,v)$ is the marginal of the joint distribution $p_t({\bf X})\equiv p_t(x,v,y)$ which obeys FP equation (\ref{EqFP1}). Specifically, for model V, 
\begin{align}
\label{EqFPV}
\partial_t  p_t({\bf X})&=-\partial_x [vp_t({\bf X})] -\frac{1}{m}\partial_v\big[-[kx+\gamma' y]p_t({\bf X})+J_t({\bf X})\big]-\frac{1}{\tau}\partial _y \big[(v-y)p_t({\bf X})-\frac{T'}{\gamma'\tau}\partial_yp_t({\bf X})\big]
\end{align}
where 
\begin{align}
\label{EqJtV}
J_t({\bf X})=-\gamma [ v+\frac{T}{m}\partial_v \ln  p_t({\bf X})] p_t({\bf X}))
\end{align}
and $T'=\gamma' S_{v_n}/2$ like before. Integrating over $y$ yields the FP equation for $p_t(x,v)$,
\begin{align}
\label{EqFPKQn4}
\partial_t  p_t(x,v)=-\partial_x [vp_t(x,v)] -\frac{1}{m}\partial_v\left[[-kx+\tilde F_{\textnormal{fb}}(x,v,t)] p_t(x,v)+ J_t(x,v)\right]
\end{align}
where the current 
\begin{align}
\label{EqJtV1}
 J_t(x,v)=\int dy\: J_t({\bf X})=-\gamma [ v+\frac{T}{m}\partial_v \ln  p_t(x,v)] p_t(x,v)
\end{align}
has the same definition as before, and the apparent feedback force is now defined as
\begin{align}
\label{EqFeffV}
\tilde F_{\textnormal{fb}}(x,v,t)=-\gamma' \tilde y(x,v,t) 
\end{align}
where
\begin{align}
\label{Eqbary}
{\tilde y}(x,v,t)=\int dy \: y \: p_t(y\vert x,v) =\frac{1}{p_t(x,v)}\int dy \: y \: p_t({\bf X})\ .
\end{align} 
Therefore, the only difference with the preceding calculations for $\tau=0$ lies in the definition of $\tilde F_{\textnormal{fb}}(x,v,t)$. By taking the time derivative of  $S_{\textnormal{sys}}(t)$, we thus again obtain Eq. (\ref{Eq2ndL3})  with $\dot {\tilde \Sigma} (t)$ and $\dot S_m(t)$ given  by Eqs. (\ref{EqPERP}) and (\ref{EqSm}), respectively, whereas the entropy pumping is now given by
\begin{subequations}
\label{EqSpux}
\begin{align}
\label{EqSpuxa}
\dot S_{\textnormal{pu}}(t)&=\frac{1}{m}\int dx dv \: p_t(x,v)\frac{\partial \tilde F_{\textnormal{fb}}(x,v,t)}{\partial v}\\
\label{EqSpuxb}
&= -\frac{\gamma'}{m} \int dx dv \: p_t(x,v)\partial_v{\tilde y}(x,v,t) \ .
\end{align}
\end{subequations}
The physical meaning of this contribution is again more transparent in the NESS where $p_{\textnormal{st}}({\bf X})$ is given by the Gaussian distribution (\ref{Eqpss}) and the effective (kinetic) temperature is
\begin{align}
\label{EqTeffV}
T_{\textnormal{eff}}^{(v)}\equiv m\sigma_{2,2}=\frac{ \gamma_{\textnormal{eff}} -\gamma'}{\gamma_{\textnormal{eff}}}T+\frac{\gamma'}{ \gamma_{\textnormal{eff}}}T'
\end{align}
with $ \gamma_{\textnormal{eff}}$ defined by Eq. (\ref{Eqgtilde}). A straightforward calculation  then yields
\begin{align}
\label{Eqystar}
\tilde y_{\textnormal{st}}(x,v)&=\frac{<xy>_{\textnormal{st}}}{<x^2>_{\textnormal{st}}}x+\frac{<vy>_{\textnormal{st}}}{<v^2>_{\textnormal{st}}}v\nonumber\\
&=\frac{\sigma_{1,3}}{\sigma_{1,1}}x+\frac{\sigma_{2,3}}{\sigma_{2,2}}v
\end{align}
where the elements of the covariance matrix ${\boldsymbol \sigma}$ are given by Eqs. (\ref{EqsigmaV}). The apparent feedback force defined by Eq. (\ref{EqFeffV}) thus includes an additional contribution proportional to the instantaneous position of the resonator. However, from Eq. (\ref{EqSpux}), only the viscous part of  $\tilde F_{\textnormal{fb,st}}(x,v)$ contributes to the entropy pumping and the rate can be again expressed as  $\dot S_{\textnormal{pu}}=-\tilde \gamma'/m$ with an apparent friction coefficient 
\begin{subequations}
\label{EqgeffV}
\begin{align}
\label{EqgeffVa}
\tilde \gamma'&\equiv \gamma' \frac{\sigma_{2,3}}{\sigma_{2,2}}\\
\label{EqgeffVb}
&=\gamma \frac{T-T_{\textnormal{eff}}^{(v)}}{T_{\textnormal{eff}}^{(v)}} 
\end{align}
\end{subequations}
where Eq. (\ref{EqrelVa}) has been used to go from Eq. (\ref{EqgeffVa}) to (\ref{EqgeffVb}) (note that $\tilde \gamma'\ne \gamma' (T_{\textnormal{eff}}^{(v)}-T')/T_{\textnormal{eff}}^{(v)}$, in contrast with Eq. (\ref{Eqgammaeffa})). Specifically, we obtain 
\begin{subequations}
\label{EqtildeSigmaxst}
\begin{align}
\label{EqtildeSigmaxsta}
\dot {\tilde \Sigma}&=\dot Q(\frac{1}{T}-\frac{1}{T_{\textnormal{eff}}^{(v)}})\\
&=\frac{\gamma \gamma'^2 }{m \gamma_{\textnormal{eff}}}\frac{(T-T')^2}{[(\gamma_{\textnormal{eff}} -\gamma')T+\gamma' T']T} \ ,
\end{align}
\end{subequations}
\begin{align}
\dot S_{m}\equiv \frac{\dot Q}{T}=-\frac{\gamma \gamma' }{m\gamma_{\textnormal{eff}}}\frac{T-T'}{T}\ ,
\end{align}
and
\begin{align}
\label{Eqspux}
\dot S_{\textnormal{pu}}&=-\frac{\gamma \gamma' }{m}\frac{T-T'}{(\gamma_{\textnormal{eff}} -\gamma')T+\gamma' T'}\ ,
\end{align}
which generalize Eqs.  (\ref{EqdSigma}), (\ref{EqdSm}), and  (\ref{Eqdspu}), respectively.  Since $\gamma_{\textnormal{eff}} \rightarrow \gamma + \gamma'$ when $\tau \rightarrow 0$, these equations are recovered in this limit, as it must be. Furthermore, the apparent EP still reaches a maximum as a function of the gain $g=\gamma'/\gamma$ when $T_{\textnormal{eff}}^{(v)}$ is minimal.

Following the same line of reasoning as for $\tau=0$, we can introduce an effective Langevin equation that reproduces the same average heat flow $\dot Q$ and  the same average EP rate $\dot {\tilde \Sigma} $ in the steady state as the original model while allowing to properly define  corresponding fluctuating quantities. The Fokker-Planck equation (\ref{EqFPKQn4}) suggests to simply replace the actual feedback force by the apparent one defined by Eq. (\ref{EqFeffV}), that is to replace $y$ by $\tilde y_{\textnormal{st}}(x,v)$ in Eq. (\ref{EqmodelVa}). From Eqs. (\ref{Eqystar}) and (\ref{EqgeffVa}), the effective  Langevin equation thus reads
\begin{align}
\label{EqLeffx}
m\ddot x +(\gamma+\tilde \gamma')\dot x+(k+\gamma' \frac{\sigma_{1,3}}{\sigma_{1,1}})x =\sqrt{2\gamma T}\: \xi(t) \ ,
\end{align}
which can be rewritten, using  Eqs. (\ref{EqrelV}), as
\begin{align}
\label{EqLeffx1}
m\ddot x +\gamma\left( \frac{T}{T_{\textnormal{eff}}^{(v)}}\right) \dot x+k \left(\frac{T_{\textnormal{eff}}^{(v)}}{T_{\textnormal{eff}}^{(x)}}\right)x =\sqrt{2\gamma T}\: \xi(t) \ 
\end{align}
where $T_{\textnormal{eff}}^{(x)}\equiv k<x^2>_{\textnormal{st}}=k\sigma_{1,1}$ is another effective temperature characterizing the motion of the Brownian entity\footnote{Note from Eqs. (\ref{EqsigmaV}) that $T_{\textnormal{eff}}^{(v)}=T_{\textnormal{eff}}^{(x)}=T$ when $T'=T$. The Brownian entity is then at equilibrium with the environment. One also has $T_{\textnormal{eff}}^{(v)}=T_{\textnormal{eff}}^{(x)}$ for $\tau=0$ so that Eq. (\ref{EqLeffx1}) gives back Eq. (\ref{EqLeff}), as it must be.} (see below section 4C). By construction,  this equation leads to the same marginal distribution $p_{\textnormal{st}}(x,v)$ as Eqs. (\ref{EqmodelV}). The apparent heat dissipated in the environment is then given by
\begin{align}
\label{EqDeltasmeff2}
\tilde q[\{x_s\}]\equiv T\Delta \tilde s_m[\{x_s\}]=-\int_0^t ds\: [m \ddot x_s+\gamma' \tilde y_{\textnormal{st}}(x_s,\dot x_s)+kx_s]\circ \dot x_s  \ ,
\end{align}
and the corresponding apparent EP functional $\tilde \sigma [\{x_s\}]\equiv  \Delta s_{\textnormal{sys}}+\Delta \tilde s_{m}[\{x_s\}]+(\tilde \gamma'/m)t$ gives back Eq. (\ref{EqtildeSigmaxst}) upon averaging. The reason for using the term ``apparent'' like in \cite{MLBBS2012} should now be clear.  Apart from the presence of the entropy pumping contribution, which is specific to the present problem, the definition of the path-dependent apparent EP is the same as the one introduced in \cite{MLBBS2012} for an overdamped dynamics\footnote{The present system, however, is quite different from the one studied in \cite{MLBBS2012} which involves two interacting Brownian particles and two NESSs that can be controlled independently. Here, there is only one Brownian entity and a single steady state.}: the total force acting on the observed particle, which depends on the ``hidden" degrees of freedom (here, $y$), is replaced by its conditional expectation, that is by its projection on the subspace spanned by the accessible degrees of freedom (here, $x$ and $v$)\footnote{$\tilde y(x,v,t)$, as defined by Eq. (\ref{Eqbary}), is in fact  the minimum mean-squared error (MMSE) estimator of $y$ for given $x$ and $v$, that is the Bayes estimator that minimizes the mean-squared error $<(\tilde y -y)^2>$, where the expectation is taken over $x,v$ and $y$. Since all variables are jointly Gaussian in the NESS, this estimator is a linear function of $x$ and $v$, as shown by Eq. (\ref{Eqystar}).}. By definition, one has $\tilde \sigma [\{x_s\}]=\ln {\cal \tilde P}_+[\{x_s\}]/{\cal \tilde P}_-[\{\hat x_s\}] \ne  \ln {\cal P}_+[\{x_s\}]/{\cal P}_-[\{\hat x_s\}]$, which implies that the apparent EP functional obeys a FT with the trajectories $\{x_s\}_{s\in[0,t]}$ generated by the effective Langevin equation instead of the actual trajectories generated by Eqs. (\ref{EqmodelV})\footnote{This is also true for the apparent EP defined in [36]. This trajectory-dependent functional  does not obey a FT if the dynamics is described
by the original Langevin equation (only FT-like symmetries may be preserved, depending on the experimental parameters). On the other hand, if one defines an effective dynamics (which generates different stochastic trajectories) by replacing the actual force acting on the observed particle by its conditional expectation computed with the full stationary probability distribution, the apparent EP then obeys a FT with this effective dynamics.} (here $\tilde P_-[\{\hat x_s\}]$ is the probability  of the time-reversed path generated by the Langevin equation conjugate to Eq. (\ref{EqLeffx})  in which the apparent friction coefficient $\tilde \gamma' $ is replaced by $-\tilde \gamma'$ and the other terms are unchanged).

Of course, a more standard picture is recovered if one considers the EP along trajectories ${\{\bf X}_s\}_{s\in[0,t]}\equiv(\{x_s\},\{y_s\})_{s\in[0,t]}$ in the ``super-system''.  As already stressed, this is not the viewpoint adopted in this work, in contrast with \cite{K2012}. Moreover, this EP is rather artificial in the present context  since Eqs. (\ref{EqmodelV}) do not describe the actual physical processes inside the controller. On the other hand, the comparison with the apparent EP defined above may be interesting from the perspective of the influence of coarse graining on entropy production, in particular in the light of the analysis carried out in \cite{CPV2012}. Indeed, if one forgets the harmonic potential $kx^2/2$, model V is identical to the two-temperatures underdamped model considered in \cite{CPV2012} (Appendix B) that describes the irreversible dynamics of a massive tracer in a granular fluid\cite{VBPV2009,PV2009,SVCP2010}.  For completeness, and also because this sheds some light on the analysis of \cite{K2012}, the calculation of the EP in the super-system (the ``total" EP) is detailed in Appendix C.
Note in particular Eq. (\ref{Eqinequality}) which states that the average total EP rate is always larger than the average apparent EP rate. This is consistent with the general expectation that an incomplete description of a system results in an underestimation of the actual dissipation. However, in the present case, since $\tilde \sigma [\{x_s\}] \ne  \ln {\cal P}_+[\{x_s\}]/{\cal P}_-[\{\hat x_s\}]$, this inequality does not follow from the general argument that a Kullback-Leibler divergence (or relative entropy) always decreases upon coarse graining\cite{CT2006} (see also \cite{GPVdB2008}).

\subsection{Model P}

Since it would be tedious to repeat everything for model P, we only point out the main differences with model V and give the main results. We first recall that the model described by Eqs. ({\ref{EqmodelP}) is ill-defined for $\tau=0$ because the measurement noise on the resonator position $x_n(t)=\sqrt{S_{x_n}}\eta (t)$ is approximated by a Gaussian white noise. This implies that some quantities diverge in the limit $\tau \rightarrow 0$ (see Eqs. (\ref{EqsigmaP})), in particular the effective kinetic temperature $T_{\textnormal{eff}}^{(v)}$ in the NESS which is given by 
\begin{align}
\label{EqTeffPv}
T_{\textnormal{eff}}^{(v)}=\frac{ \gamma_{\textnormal{eff}} -\gamma'}{ \gamma_{\textnormal{eff}}} T + \frac{\gamma' }{\gamma} \frac{\gamma_{\textnormal{eff}} -\gamma-\gamma' }{ \gamma_{\textnormal{eff}}} T'
\end{align}
where $T'\equiv \gamma'S_{x_n}/(2\tau^2)$ has the dimension of a  temperature (the important new  feature is that $T'$ diverges  for $\tau \rightarrow 0$, in contrast with the corresponding quantity in model V). On the other hand, the other effective temperature $T_{\textnormal{eff}}^{(x)}$, which is  the one usually considered in experiments\cite{PDMR2007}, remains finite in this limit\footnote{Here, to facilitate the comparison with  experiments, we use the parameters $\tau_0=m/\gamma$, $\omega_0=\sqrt{k/m}$ and $Q_0=\sqrt{m k}/\gamma$ to describe the resonator instead of $k$, $m$, and $\gamma$. The temperature $T_{\textnormal{eff}}^{(x)}$ is denoted $T_{\textnormal{mode}}$ in \cite{PDMR2007}.},
 \begin{align}
\label{EqTeffPx}
T_{\textnormal{eff}}^{(x)}&=\frac{ \gamma_{\textnormal{eff}} -\gamma' (1+ \frac{\tau}{\tau_0})}{ \gamma_{\textnormal{eff}}}T+(Q_0\frac{\tau}{\tau_0})^2\frac{\gamma'}{ \gamma_{\textnormal{eff}}} T'\nonumber\\
&\rightarrow \frac{T}{1 +g}+\frac{k\omega_0}{2Q_0}(\frac{g^2}{1+g})S_{x_n} \ ,
%&\rightarrow \frac{\gamma}{\gamma +\gamma'}T+\frac{k}{2m}\frac{\gamma'^2}{\gamma +\gamma'}S_{x_n} \nonumber\\
\end{align}
which is Eq. (5) in \cite{PDMR2007} (with $k_B=1$ and the two-sided convention for the spectral densities, see footnote [29] in Appendix A). As pointed out in Appendix A (see Eqs. (\ref{EqPSDP})), this can be traced back to the behavior of  $S_{xx}(\omega)$, the power spectral density (PSD) of $x$, at large frequencies: in the limit  $\tau \rightarrow 0$, the integral of $S_{xx}(\omega)$ over $\omega$ is finite whereas the integral of $S_{vv}(\omega)=\omega^2 S_{xx}(\omega)$ diverges, which is also the case for the integral of $S_{x'x'}(\omega)$,  the PSD of the measured displacement $x'=x+x_n$ (i.e. the PSD of the detector output)\footnote{\label{toto}In the analysis of the experimental spectra, the integration over $\omega$ is actually performed in a limited band around the resonance frequency $\omega_0$ (see for instance the discussion in \cite{A2009} for the LIGO interferometer). In model P, $\tau$ sets the minimal accessible time scale and in principle the upper limit of the integrals must be of the order of $2\pi/\tau$. This limit can be extended to $+\infty$ if the integrals converge.}. 

 Another significant difference with model V is the fact that there is no positive value of the feedback gain $g=\gamma'/\gamma$ for which the resonator is at equilibrium with the environment. This can be readily seen from the above equations which show that $T_{\textnormal{eff}}^{(v)}$ and $T_{\textnormal{eff}}^{(x)}$ cannot be simultaneously equal to the heat bath temperature $T$. In other words, the detailed balance condition is never satisfied and there is always dissipation in the system\footnote{It does exist a value of $g$ for which $T_{\textnormal{eff}}^{(x)}=T$, but the corresponding kinetic temperature $T_{\textnormal{eff}}^{(v)}$ is then larger than $T$. In practice, this value of $g$ is very large (see Fig. 3(a) for instance) and  this situation has not been observed experimentally to the best of our knowledge.}.

Using the same method as previously, we first define the apparent EP at the ensemble level. The Fokker-Planck is somewhat more complicated than in model V and reads 
\begin{align}
\label{EqFP20}
\partial_t p_t({\bf X})&=-\partial_x[vp_t({\bf X})]+\frac{1}{m}\partial_v\left[[(k+\frac{\gamma'}{\tau})x+\gamma v-\frac{\gamma'}{\tau}y]p_t({\bf X})\right]+\frac{1}{\tau}\partial_y[(y-x)p_t({\bf X})]+\frac{\gamma T+ \gamma' T'}{m^2}\frac{\partial^2}{\partial v^2}p_t({\bf X})\nonumber\\
& +\frac{T'}{\gamma'}\frac{\partial^2}{\partial y^2}p_t({\bf X})-\frac{2T'}{m}\frac{\partial^2}{\partial v\partial y}p_t({\bf X})\ ,
\end{align}
the cross derivatives arising from the fact that the noises in the r.h.s of Eqs. (\ref{EqmodelPa}) and (\ref{EqmodelPb}) are correlated. However, these terms to do contribute after integrating over $y$, and the FP equation for the marginal probability distribution $p_t(x,v)=\int dy \: p_t(x,v,y)$ can be  written as
\begin{align}
\label{EqFPKQn5}
\partial_t  p_t(x,v)=-\partial_x [vp_t(x,v)] -\frac{1}{m}\partial_v\left[[-(k+\frac{\gamma'}{\tau})x+\tilde F_{\textnormal{fb}}(x,v,t)]p_t(x,v) +J_t(x,v)\right]
\end{align}
where the current $J_t(x,v)$  has the same expression as in model V (see Eq. (\ref{EqJtV1})), and the apparent feedback force is now defined as
\begin{align}
\tilde F_{\textnormal{fb}}(x,v,t)=\frac{\gamma'}{\tau}[\tilde y(x,v,t) -\frac{\tau T'}{m}\partial_v\ln p_t(x,v)]
\end{align}
with $\tilde y(x,v,t) $ given by Eq. (\ref{Eqbary}). This is the main difference with model V, and  $\dot {\tilde \Sigma} (t)$ and $\dot S_m(t)$ are again given  by Eqs.  (\ref{EqPERP}) and (\ref{EqSm}) in the entropy balance equation (\ref{Eq2ndL3}), whereas the entropy  pumping is
\begin{subequations}
\label{EqSpuxP}
\begin{align}
\label{EqSpuxPa}
\dot S_{\textnormal{pu}}(t)&=\frac{1}{m}\int dx dv \: p_t(x,v)\partial_v \tilde F_{\textnormal{fb}}(x,v,t)\\
\label{EqSpuxPb}
&= \frac{\gamma'}{\tau m} \int dx dv \: p_t(x,v)\partial_v [\tilde y_t(x,v)-\frac{\tau T'}{m}\partial_v \ln p_t(x,v)] \ .
\end{align}
\end{subequations}
Eq. (\ref{EqSpuxPb}) is in general different from Eq. (\ref{EqSpuxb}), but  the entropy pumping rate in the NESS, after using Eq. (\ref{Eqystar}) and Eq. (\ref{EqrelPb}), can be again expressed as $\dot S_{\textnormal{pu}}=-\tilde \gamma'/m$ with an apparent friction coefficient $\tilde \gamma'$ given by Eq. (\ref{EqgeffVb}). Similarly,  $\dot {\tilde \Sigma}$ and $\dot S_m$ are given by  Eq. (\ref{EqdotSigmaa}) and Eq. (\ref{EqdotSma}), respectively.  Hence, all the terms in the entropy balance equation are formally the same as in model V and only the value of the effective kinetic temperature changes. Explicitly, we obtain
\begin{align}
\label{EqeprP}
\dot {\tilde \Sigma}&=\dot Q(\frac{1}{T}-\frac{1}{T_{\textnormal{eff}}^{(v)}})\nonumber\\
&= \frac{\gamma'^2}{m\gamma_{\textnormal{eff}} } \frac{[\gamma T-(\gamma_{\textnormal{eff}} -\gamma -\gamma')T']^2}{[\gamma (\gamma_{\textnormal{eff}} -\gamma')T+\gamma' (\gamma_{\textnormal{eff}} -\gamma- \gamma')T']T} \ ,
\end{align}
\begin{align}
\label{EqEPSmP}
\dot S_m&\equiv \frac{\dot Q}{T}=-\frac{\gamma'}{m\gamma_{\textnormal{eff}}} \frac{\gamma T-(\gamma_{\textnormal{eff}} -\gamma -\gamma')T'}{T}\ ,
\end{align}
and 
\begin{align}
\label{Eqspuy}
\dot S_{\textnormal{pu}}=- \frac{\gamma \gamma'}{m} \frac{\gamma T-(\gamma_{\textnormal{eff}} -\gamma -\gamma')T'}{\gamma (\gamma_{\textnormal{eff}} -\gamma')T+\gamma' (\gamma_{\textnormal{eff}} -\gamma- \gamma')T'}
\end{align}
with $\gamma_{\textnormal{eff}}$ given by Eq. (\ref{Eqgtilde}).  As it must be, the results of \cite{KQ2004} recalled in section 3 are recovered by setting $T'=0$ (i.e. $S_{x_n}=0$) and $\tau=0$. On the other hand, if $S_{x_n}\ne 0$, both $\dot {\tilde \Sigma}$ and $\dot S_m$ diverge like $1/\tau$ as $\tau \rightarrow 0$ (whereas the entropy pumping rate stays finite), as a consequence of the divergence of the noise temperature $T'$. 

To derive an effective Langevin equation in the NESS, we cannot simply replace $\dot y$ by $\dot {\tilde y}_{\textnormal{st}}(x,\dot x)$ in Eq. (\ref{EqmodelP0a})  as this would introduce an effective mass in the problem. What must be done is to project both $y$ and the noise $\eta$ on the subspace spanned by the  variables $x$ and $v$. We thus define 
\begin{align}
\tilde \eta_{\textnormal{st}}(x,v)&=\frac{<\eta \circ x>}{<x^2>_{\textnormal{st}}}x+\frac{<\eta \circ v>}{<v^2>_{\textnormal{st}}}v\nonumber\\ 
&=- \frac{1}{2T_{\textnormal{eff}}^{(v)}}\sqrt{2\gamma' T'}\: v \ ,
\end{align}
where we have used $<\eta \circ x>=0$ and $<\eta \circ v>=-1/(2m)\sqrt{2\gamma' T'}$ to derive the second equality. Then, replacing $y$ by $\tilde y_{\textnormal{st}}(x,\dot x)$ and $\eta$ by $\tilde \eta_{\textnormal{st}}(x,\dot x)$ into Eq. (\ref{EqmodelPa})  yields  the effective Langevin equation
\begin{align}
\label{EqLeffP}
m\ddot x+(\gamma +\tilde \gamma')\dot x+(k+\frac{\gamma'}{\tau}-\frac{\gamma'}{\tau}\frac{\sigma_{1,3}}{\sigma_{1,1}}) x=\sqrt{2\gamma T} \xi(t)\ ,
\end{align}
which can be exactly rewritten, using Eqs. (\ref{Eqystar}), (\ref{EqrelPb}), and (\ref{EqrelPc}),  as Eq. (\ref{EqLeffx1}) in model V\footnote{This is not surprising since the effective Langevin equation, by construction, must yield the same marginal probability distribution $p_{\textnormal{st}}(x,v)$  as Eqs. (\ref{EqmodelP}), and this quantity has formally the same expression  in models V and  P, and only the effective temperatures $T_{\textnormal{eff}}^{(v)}$ and $T_{\textnormal{eff}}^{(x)}$ are different.}. The apparent heat dissipated in the environment is  defined as
\begin{align}
\tilde q[\{x_s\}]\equiv T\Delta \tilde s_m[\{x_s\}]=-\int_0^t ds\: [m \ddot x_s+\tilde \gamma'\dot x_s+(k+\frac{\gamma'}{\tau}-\frac{\gamma'}{\tau}\frac{\sigma_{1,3}}{\sigma_{1,1}}) x_s]\circ \dot x_s  \ ,
\end{align}
and the corresponding apparent EP functional $\tilde \sigma [\{x_s\}]\equiv  \Delta s_{\textnormal{sys}}+\Delta \tilde s_{m}[\{x_s\}]+(\tilde \gamma'/m)t$ obeys fluctuation theorems with the trajectories generated by Eq. (\ref{EqLeffP}) and the conjugate equation (where $\tilde \gamma'$ is replaced by $-\tilde \gamma'$ but $\gamma'$  in the coefficient of $x$ is  not changed), while giving back Eq. (\ref{EqeprP}) upon averaging.

Finally, we can again compare the apparent EP to the  (total) EP in the super-system that contains the full statistical information on the degrees of freedom $x$ and $y$. The calculation is more complicated that the one performed in Appendix C for model V because the noises coming into play in Eqs. (\ref{EqmodelP}) are correlated. Here, for brevity, we only give the expression of the entropy balance equation in the NESS,
\begin{align}
\label{EqeptotP}
\dot \Sigma&=\dot S_m+\dot S_{m'}\nonumber\\
&=\frac{\gamma}{m}\frac{T_{\textnormal{eff}}^{(v)}-T}{T}+\frac{1}{\tau}\frac{T_{\textnormal{eff}}^{(v)}-T_{\textnormal{eff}}^{(x)}}{T'}\nonumber\\
&=\frac{\gamma'}{m\gamma_{\textnormal{eff}}} \frac{(\gamma T+\gamma' T')T+(\gamma_{\textnormal{eff}} -\gamma- \gamma')T'^2}{TT'}
\end{align}
where Eqs. (\ref{EqrelP}) are used to obtain the last expression. This quantity does not vanish for $T'=T$ and it can be easily checked that it is always larger than $\dot {\tilde \Sigma}$. We also notice  that $T\dot S_m +T' \dot S_{m'}=(\gamma'/m)T'$. Therefore, if one associates $T' \dot S_{m'}$ to the heat flow $\dot Q'$ from a second reservoir at temperature $T'$, one has $\dot Q+\dot Q'\ne 0$. Accordingly, $\dot \Sigma \ne \dot Q (1/T-1/T')$, in contrast with model V. 

\begin{figure}[hbt]
\begin{center}
\includegraphics[width=10cm]{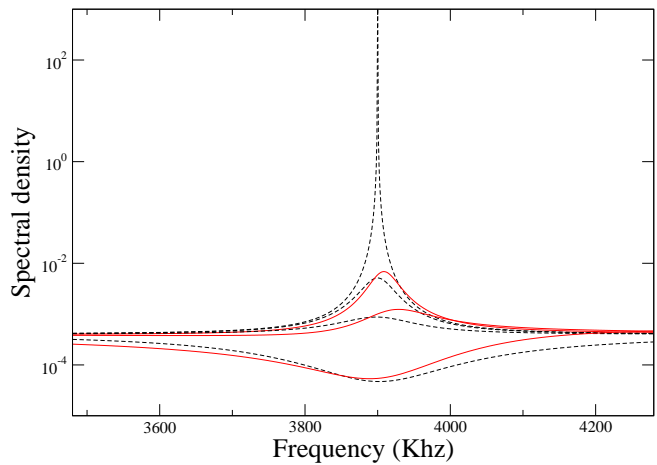}
\caption{\label{Fig2}  Spectral density $S_{x'x'}(\omega)$ (in $\AA^2/$Hz) given by Eq. (\ref{EqPSDPb}) for different values of the feedback gain $g$. The resonator parameters correspond to the cantilever 1 studied in \cite{PDMR2007} (see text). The cantilever is cooled from a base temperature $T=4.2 K$. From top to bottom: $g=0, 544, 1321,5690$. Black dashed lines: $\tau=0$; red solid lines: $\tau=0.1$ms. }
\end{center}
\end{figure}

\begin{figure}[hbt]
\begin{center}
\includegraphics[width=12cm]{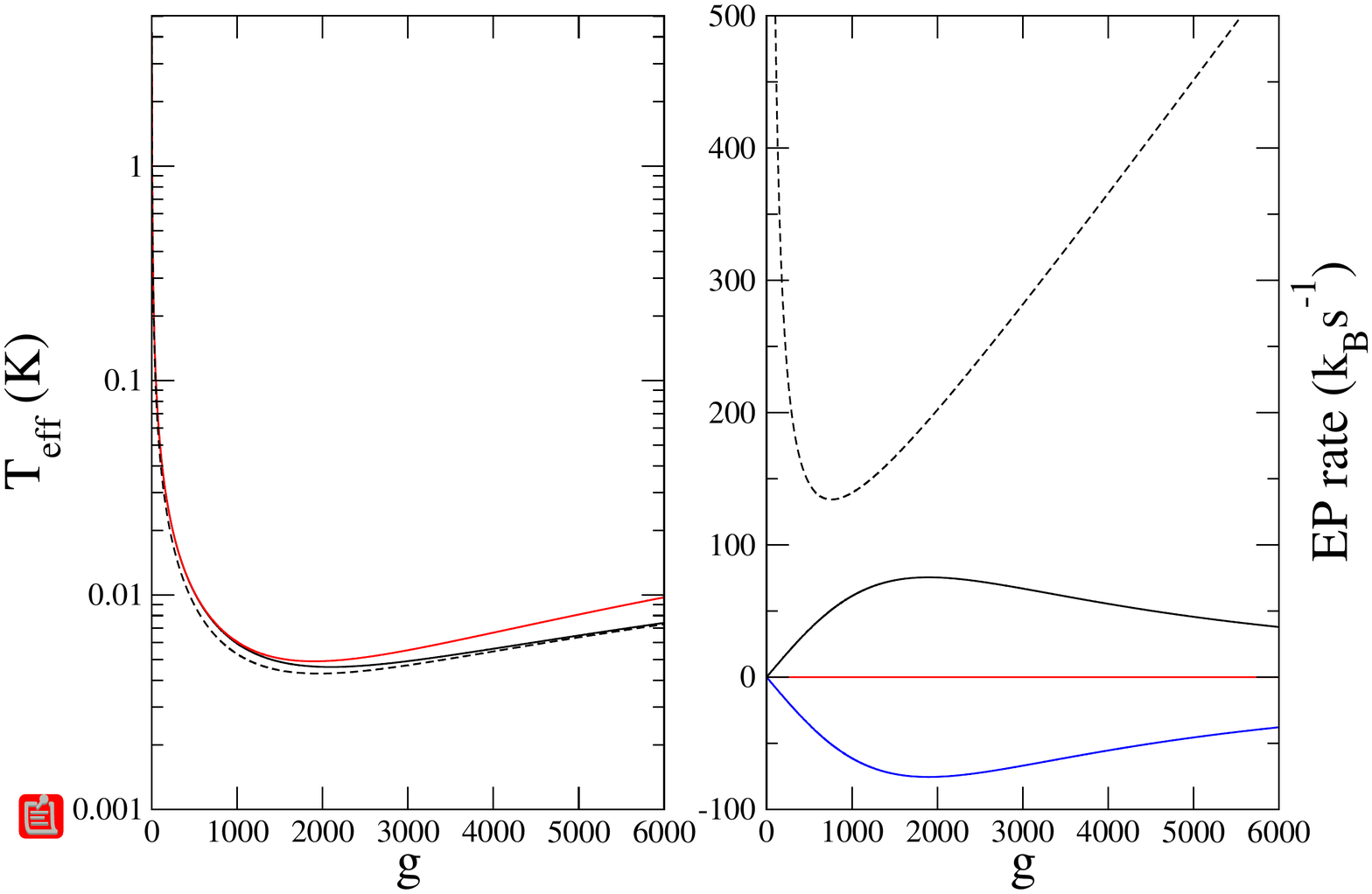}
\caption{\label{Fig2} (a) Resonator temperature $T_{\textnormal{eff}}^{(x)}$ as a function of the gain $g$ for $\tau=0$ (black dashed line) and $\tau=0.1$ms (black solid line). The red line is the kinetic temperature $T_{\textnormal{eff}}^{(v)}$ for $\tau=0.1$ms. (b) Different contributions to the entropy production rate for $\tau=0.1$ms according to Eqs. (\ref{EqeprP}), (\ref{EqEPSmP}), and (\ref{Eqspuy}) : $\dot {\tilde \Sigma}$ (black solid line), $\dot S_m$ (red solid line), $\dot S_{\textnormal{pu}}$ (blue solid line). The dashed line is the total entropy production rate $\dot \Sigma$ in the super-system given by Eq. (\ref{EqeptotP}).}
\end{center}
\end{figure}
To illustrate the above equations, let us consider the feedback cooling of the fundamental mechanical mode of a cantilever. As an example, we take the ultrasoft silicon cantilever studied in \cite{PDMR2007}, with intrinsic quality factor $Q_0=44200$, resonant frequency $\omega_0=3.9$KHz, and  spring constant $k=86 \mu N/m$.   This cantilever is cooled from a base temperature of $4.2$ K, and  the spectral density of the measurement noise  is $\sqrt{S_{x_n}}\approx 10^{-2}\AA/ \sqrt{\mbox{Hz}}$, as estimated from fits of the measured spectra. Specifically, we take $S_{x_n}=4.10^{-4} \AA^2/ \mbox{Hz}$, which from Eq. (\ref{EqTeffPx}) yields $T_{\textnormal{eff}}^{(x)}(g=544)=8.3$ mK for $\tau=0$ in agreement with the value indicated in Fig. 3 of \cite{PDMR2007}. Finally, we choose the value $\tau=0.1$ms for the relaxation time of the feedback mechanism. This choice is rather arbitrary and is mainly done to illustrate the model behavior (the actual relaxation time in the experiment described in [3] is certainly much smaller: see the discussion at the end of the section). Note that this value is much smaller than the effective momentum relaxation time $\tau_0/(1+g)$, even for the largest value of $g$ considered in the experiments (and moreover $2\pi/\tau\gg \omega_0$). This guarantees that the resonator is still efficiently cooled in the vicinity of its resonant frequency. 

As shown in Fig. 2, this small but finite value of $\tau$ slightly modifies the measured spectral density $S_{x'x'}(\omega)$ of the cantilever computed from Eq. (\ref{EqPSDPb}). In particular, the resonant frequency is now dependent on the feedback gain\footnote{The shape of the PSD may significantly change if $\tau$ is large, especially if the quality factor of the oscillator is small. Such effects are discussed in \cite{GRBC2009}. Since this is not the main purpose of our  work, we have not performed a systematic study of the model behavior as a function of the different parameters.}. 
Note that the peak in $S_{{x'x'}}(\omega)$ changes into a dip in the high gain regime as the detector noise sent back to the Brownian system dominates and acts to heat the mechanical device, an effect known as ``noise squashing"\cite{PCBH2000,PDMR2007,VBMF2010}. 
Accordingly, integrating the observed spectrum $S_{x'x'}(\omega)$ over $\omega$ leads to an underestimation of the actual resonator temperature. The true resonator motion $S_{xx}(\omega)$ can be recovered by using the theoretical expression, Eq. (\ref{EqPSDPa}), or can be directly measured by adding a second transducer outside the feedback loop\cite{L2010,MMSPP2012}. 

As could be expected, in presence of a finite relaxation time, the feedback cooling becomes less effective, and the minimal achievable temperature $T_{\textnormal {eff}}^{(x)}$ increases, as shown in Fig. 3 (a), albeit very slightly (from $4.3$ mK for $\tau=0$ to $4.6$ mK  for $\tau=0.1$ms). On the other hand, for this value of $\tau$, the two temperatures $T_{\textnormal{eff}}^{(x)}$ and $T_{\textnormal{eff}}^{(v)}$ are very close one to each other around the minimum. Accordingly,  the apparent EP rate $\dot {\tilde \Sigma}$ displayed in Fig. 3(b) is maximal when $T_{\textnormal{eff}}^{(x)}$ is close to its minimum. We also notice that the essential contribution to $\dot {\tilde \Sigma}$ comes from the entropy pumping since $\dot S_m$ is very small in the whole range of $g$ that is experimentally explored.  This was also observed in model V for $\tau=0$ (see Fig. 1). This picture changes for much larger values of the gain (typically $g > 10^6$) as the resonator is heated instead of cooled. The main contribution to  $\dot {\tilde \Sigma}$ then comes  from the heat dissipated into the environment and  entropy pumping is  negligible. $\dot {\tilde \Sigma}$ thus reaches a minimum for some large value of $g$ (beyond the scale of Fig. 3 (b) and unrelated to the minimum observed  for the total EP rate $\dot {\Sigma}$), but in contrast with model V, this minimum is positive since there is always dissipation in the system, as we already stressed.

Finally, let us again emphasize that the feedback relaxation time $\tau$ is expected to play a role only for resonance frequencies in the MHz range onwards. In this respect, the value $\tau=0.1$ ms chosen here for illustrative purposes only is certainly too large. Taking a much smaller value would drastically change the qualitative picture in Fig. 3 since $\dot {\tilde \Sigma}$ and $\dot S_m$ both diverge as $\tau \rightarrow 0$ (whereas $\dot S_{\textnormal{pu}}$ is finite), as already noticed. This spurious behavior of the model comes from the (commonly made) assumption that the measurement noise is white. Introducing some cutoff at large frequencies would suppress these divergences and allow us to take a more realistic value of $\tau$. This, however, would add another (rather arbitrary) parameter in the model and somewhat complicate the description. 

\section{Conclusion}

In this paper, we have generalized the results of \cite{KQ2004,KQ2007,MR2012} to take into account the effect of measurement errors (or detector noise) on the entropy production (EP) in a cold damping process. This has led us to consider two models (called P and V) that distinguish whether the position of the resonator or its velocity (in practice an electric current) is the observable, as the two situations actually occur in experimental setups. We also have assigned a finite relaxation time to the feedback mechanism, which may play a role at high frequencies, but is also required to regularize model P when the detector noise is  white. This makes the feedback control non-Markovian. 

To define the EP, we have adopted the viewpoint of the controlled system, as in \cite{KQ2004,KQ2007,MR2012} and in most recent studies of the thermodynamic behavior of feedback-controlled systems. In this framework, we have defined and computed in the nonequilibrium steady state the {\it entropy pumping} that describes the entropy reduction in the system due to its interaction with the external agent which manipulates the feedback control (and which in this case is not a genuine Maxwell's demon). For error-free measurements, the entropy pumping can be ascribed to the momentum phase space contraction induced by the additional damping force. The situation is more complicated in the presence of noise as
one cannot any more relate the heat dissipated along a stochastic trajectory of the system to time irreversibility (more precisely, this identification would lead to an unphysical result, independent of the measurement noise). A proper relationship only exists if one considers the super-system that also includes the external agent, as done in \cite{K2012}.  Accordingly, in the presence of measurement errors, the entropy pumping cannot be simply associated to a contraction of  momentum phase space. This has led us to define the entropy pumping rate and the non-negative EP rate at the ensemble level by using the (coarse-grained) Fokker-Planck equation to derive the average entropy balance equation (i.e., the generalized second law). This is the proper generalization of the results of \cite{KQ2004}. In particular, the EP rate (called the {\it apparent} EP rate as in \cite{MLBBS2012}) in the nonequilibrium steady state remains finite in the limit of error-free measurements whereas the total EP rate in the super-system diverges. Moreover, we have shown that the behavior of the apparent EP as a function of the feedback gain is consistent with the expected behavior of the dissipation in a cold damping setup: it is maximal when the feedback cooling is the most efficient. In the cooling regime, it is found that the main contribution to the EP comes from the entropy pumping whereas the average heat flow coming from the bath plays a negligible role. Measurement errors decreases the entropy pumping  (in absolute value), and the dissipation is in turn reduced. It would be interesting to generalize these observations to transient regimes, for instance when  the cooling is abruptly switched on or off, as considered in\cite{PCBH2000}.

Finally, we have shown that trajectory-dependent functionals can be defined in the nonequilibrium steady state by replacing the original  Langevin dynamics by an effective dynamics. The so-defined apparent EP functional then obeys fluctuations  theorems with the new trajectories generated by this effective dynamics.

\acknowledgments

M. L. R. is grateful to T. Sagawa and S. Ito for interesting discussions about the thermodynamics of information exchanges in nonequilibrium small systems.

\appendix

\section{Stationary probability distributions and power spectral densities (PSD) for models V and P}

In this Appendix we compute the stationary probability distributions and the power spectral densities  in models V and P. These quantities are easily obtained  by noting that  Eqs. (\ref{EqmodelV}) and (\ref{EqmodelP}) describe multivariate Ornstein-Uhlenbeck processes so that standard textbooks expressions can be used.  To this aim, we rewrite these equations in the form
\begin{align}
\label{EqL3}
\frac{d{\bf X}(t)}{dt}=-{\bf \Gamma} {\bf X} (t)+{\bf \Phi}(t)
\end{align}
where ${\bf X}$ is the 3-dimensional vector $[x,v,y]$, $\bf \Gamma$ is a $3 \times 3$ damping matrix, and  ${\bf \Phi}$ is a 3-variate Gaussian process with zero mean and symmetric covariance matrix $<\phi_i(t)\phi_j(t')>=2D_{ij} \delta (t-t')$. The explicit expressions of $\bf \Gamma$ and $\bf D$ are given below. The  corresponding Fokker-Planck equation then reads
\begin{align}
\label{EqFP1}
\partial_t p_t({\bf X})=-\sum_i\partial_{x_i} J_t^{(i)}({\bf X})=\sum_{i,j}\partial_{x_i}[\Gamma_{ij}x_jp_t({\bf X}) +D_{ij}\partial_{x_j}p_t({\bf X})] \ ,
\end{align}
and the stationary solution is given by the Gaussian distribution\cite{R1989,G2003} 
\begin{align}
\label{Eqpss}
p_{\textnormal{st}}({\bf X})=(2\pi)^{-3/2}[{\mbox Det}\: {\boldsymbol \sigma}]^{-1/2} \exp ( -\frac{1}{2} {\bf X}{\boldsymbol \sigma}^{-1} {\bf X}) \ , 
\end{align} 
where the  covariance matrix ${\boldsymbol \sigma}$ is solution of the algebraic matrix equation
\begin{align}
\label{EqsigmaPV}
2{\bf D}&={\bf \Gamma  \boldsymbol\sigma}+{\bf \Gamma \boldsymbol \sigma}^T={\bf \Gamma \boldsymbol \sigma}+{ \boldsymbol \sigma \bf \Gamma}^T \ .
\end{align} 
In model V, the  $3 \times 3$ damping matrix $\bf \Gamma$ and the covariance matrix ${\bf D}$ are 
\[
{\bf \Gamma}=\left(
\begin{array}{ccc}
  0&-1& 0 \\
 k/m&\gamma/m& \gamma' /m  \\
 0&-1/\tau &1/\tau
\end{array}
\right)
\] 
\[
{\bf D}=\left(
\begin{array}{ccc}
  0&0& 0 \\
 0&\gamma T/m^2& 0 \\
 0&0&T'/(\tau^2 \gamma ')
\end{array}
\right) 
\] 
where $T'\equiv \gamma' S_{v_n}/2$. By solving Eq. (\ref{EqsigmaPV}) we then obtain the following expressions for the elements of the covariance matrix ${\boldsymbol \sigma}$:
\begin{align}
\label{EqsigmaV}
<x^2>_{\textnormal{st}}&\equiv\sigma_{1,1}=\frac{\gamma_{\textnormal{eff}}-\gamma' (1+\frac{\tau}{\tau_0})}{\gamma_{\textnormal{eff}}}\frac{T}{k}+\frac{\gamma' (1+\frac{\tau}{\tau_0})}{\gamma_{\textnormal{eff}}}\frac{T'}{k}\nonumber\\
<v^2>_{\textnormal{st}}&\equiv\sigma_{2,2}=\frac{\gamma_{\textnormal{eff}}-\gamma'}{\gamma_{\textnormal{eff}}}\frac{T}{m}+\frac{\gamma'}{\gamma_{\textnormal{eff}}}\frac{T'}{m}\nonumber\\
 <y^2>_{\textnormal{st}}&\equiv\sigma_{3,3}=\frac{\gamma}{\gamma_{\textnormal{eff}}}\frac{T}{m}+\frac{\gamma}{\gamma_{\textnormal{eff}}}\frac{\gamma_{\textnormal{eff}} - \gamma' \frac{\tau}{\tau_0}}{ \gamma' \frac{\tau}{\tau_0}}\frac{T'}{m}\nonumber\\
 <xv>_{\textnormal{st}}&\equiv\sigma_{1,2}=\sigma_{2,1}=0\nonumber\\
 <xy>_{\textnormal{st}}&\equiv\sigma_{1,3}=\sigma_{3,1}=\frac{\tau}{m} \frac{\gamma}{\gamma_{\textnormal{eff}}}(T-T')\nonumber\\
 <vy>_{\textnormal{st}}&\equiv\sigma_{2,3}=\sigma_{3,2}=\frac{1}{m} \frac{\gamma}{\gamma_{\textnormal{eff}}}(T-T')\nonumber\\
\end{align} 
where 
\begin{align}
\label{Eqgtilde}
\gamma_{\textnormal{eff}}&\equiv (\gamma+\gamma')(1+\frac{\tau}{\tau_0}) +\frac{k \tau^2}{\tau_0}\nonumber\\
&=(\gamma+\gamma')(1+\frac{\tau}{\tau_0})+\gamma (Q_0\frac{\tau}{\tau_0})^2
\end{align} 
is an effective friction coefficient (recall that $\tau_0=m/\gamma$ and $Q_0=\sqrt{mk}/\gamma=\omega_0\tau_0$). Note the useful relations\footnote{Note also that the most probable value of $y$ for a given value of $v$ is {\it not} $v$, which is  a consequence of the non-Markovian character of the feedback control.} 
\begin{subequations}
\label{EqrelV}
\begin{align}
\label{EqrelVa}
\gamma \sigma_{2,2}+\gamma' \sigma_{2,3}&= \frac{\gamma}{m}T\\
\label{EqrelVb}
m\sigma_{2,2}-k\sigma_{1,1} &=\gamma' \sigma_{1,3}\ .
\end{align} 
\end{subequations}

Similarly in model P,
 \[
{\bf \Gamma}=\left(
\begin{array}{ccc}
  0&-1& 0 \\
 (k+\gamma'/\tau)/m&\gamma/m& -\gamma'/m   \\
 -1/\tau&0 &1/\tau
\end{array}
\right)\ ,
\] 
\[
{\bf D}=\left(
\begin{array}{ccc}
  0&0& 0 \\
 0&(\gamma T+\gamma' T')/m^2& -T'/m \\
 0&-T'/m&T'/\gamma' 
\end{array}
\right) \ ,
\] 
where $T'\equiv (\gamma'/\tau^2) S_{x_n}/2$. This  yields 
\begin{align}
\label{EqsigmaP}
<x^2>_{\textnormal{st}}&\equiv \sigma_{1,1}=\frac{\gamma_{\textnormal{eff}} -\gamma' (1+ \frac{\tau }{\tau_0})}{\gamma_{\textnormal{eff}}}\frac{T}{k}+(Q_0\frac{\tau}{\tau_0})^2\frac{\gamma'}{\gamma_{\textnormal{eff}}}\frac{T'}{k}\nonumber\\
 <v^2>_{\textnormal{st}}&\equiv\sigma_{2,2}=\frac{\gamma_{\textnormal{eff}} -\gamma'}{\gamma_{\textnormal{eff}}}\frac{T}{m}+\frac{\gamma'}{\gamma}\frac{\gamma_{\textnormal{eff}} -\gamma-\gamma'}{ \gamma_{\textnormal{eff}}}\frac{T'}{m}\nonumber\\
 <y^2>_{\textnormal{st}}&\equiv\sigma_{3,3}=\frac{\gamma}{\gamma_{\textnormal{eff}}}(1+ \frac{\tau}{\tau_0})\frac{T}{k}+\frac{\tau}{\tau_0} Q_0^2\frac{\gamma}{\gamma'}\frac{\gamma_{\textnormal{eff}} - \gamma' \frac{\tau}{\tau_0}}{\gamma_{\textnormal{eff}}}\frac{T'}{k} \nonumber\\
 <xv>_{\textnormal{st}}&\equiv \sigma_{1,2}=\sigma_{2,1}=0\nonumber\\
 <xy>_{\textnormal{st}}&\equiv \sigma_{1,3}=\sigma_{3,1}=\frac{\gamma}{\gamma_{\textnormal{eff}}}(1+\frac{\tau }{\tau_0})\frac{T}{k}-(Q_0\frac{\tau}{\tau_0})^2\frac{\gamma}{\gamma_{\textnormal{eff}}}\frac{T'}{k}\nonumber\\
 <vy>_{\textnormal{st}}&\equiv\sigma_{2,3}=\sigma_{3,2}=- \frac{\tau}{\tau_0}(\frac{T}{\gamma_{\textnormal{eff}}}+ \frac{\gamma+\gamma'}{\gamma}\frac{T'}{\gamma_{\textnormal{eff}}} )\ .
\end{align} 

Note again the useful relations 
\begin{subequations}
\label{EqrelP}
\begin{align}
\label{EqrelPa}
\gamma \sigma_{1,1}+\gamma' \sigma_{1,3}&=\frac{\gamma T}{k} \\
\label{EqrelPb}
\gamma \sigma_{2,2}-\frac{\gamma'}{\tau} \sigma_{2,3}&=\gamma\frac{ T}{m}+\gamma'\frac{T'}{m}\\
\label{EqrelPc}
(k+\frac{\gamma'}{\tau})\sigma_{1,1}-\frac{\gamma'}{\tau}\sigma_{1,3}&=m\sigma_{2,2}\ .
\end{align} 
\end{subequations}
The power spectral densities  of $v$ (resp. $x$), the actual velocity (resp. displacement) of the resonator, and $v'=v+v_n$ (resp. $x'=x+x_n$), the observed velocity (resp. displacement) are obtained by using the  expression for the spectrum matrix of a multivariate Ornstein-Uhlenbeck process in the stationary state\cite{G2003},
\begin{align}
\label{Eqspectrum}
{\bf S}(\omega)=({\bf \Gamma}+i\omega {\bf 1})^{-1}(2{\bf D} )({\bf \Gamma}^T-i\omega {\bf 1})^{-1} \ .
\end{align}
For model V, this yields 
\begin{subequations}
\label{EqPSDV}
\begin{align}
\label{EqPSDVa}
S_{vv}(\omega)&=\left[ \frac{(1 +\tau^2\omega^2)\omega^2/m^2}{\vert D(\omega)\vert^2}\right]S_{F_{\textnormal{th}}}+\left[\frac{g^2\omega^2/\tau_0^2}{\vert D(\omega)\vert^2}\right]S_{v_n}\\
\label{EqPSDVb}
S_{v'v'}(\omega)&=(1 +\tau^2\omega^2)S_{y,y}(\omega)\nonumber\\
&=\left[ \frac{(1 +\tau^2\omega^2) \omega^2/m^2}{\vert D(\omega)\vert^2}\right]S_{F_{\textnormal{th}}}+ \left[\frac{(1 +\tau^2\omega^2)[(\omega_0^2-\omega^2)^2+\omega^2/\tau_0^2]}{\vert D(\omega)\vert^2}\right]  S_{v_n}
%S_{v,y}(\omega)&=\left[-\frac{(1+i\omega \tau)\omega^2/m^2}{\vert D(\omega)\vert^2}\right]S_{F_{\textnormal{th}}}+\left[g (\omega/\tau_0)\frac{\omega/\tau_0+i(\omega_0^2-\omega^2)}{\vert D(\omega)\vert^2}\right]  S_{v_n}  \ .
\end{align}
\end{subequations}
where 
\begin{align}
\label{EqDomega}
D(\omega)=[\omega_0^2-(1+\frac{\tau}{\tau_0})\omega^2]+i\omega[\frac{1+g}{\tau_0} +\tau (\omega_0^2-\omega^2)] 
\end{align}
and $S_{F_{\textnormal{th}}}\equiv 2\gamma T$ is the white spectral density of the thermal noise force\footnote{\label{convPSD} In this work we use the two-sided convention for a spectral density, {\it i.e.} $S_{\alpha,\beta}(\omega)\equiv \int_{-\infty}^{+\infty} e^{i\omega t}\phi_{\alpha,\beta}(t) dt$ where $\phi_{\alpha,\beta}(t)$ is a time-translational invariant correlation function in the stationary state.  Hence, $S_{F_{\textnormal{th}}}=2\gamma T$ for the PSD of the Langevin thermal force.  On the other hand, the one-sided convention is often used in experimental papers, for instance in \cite{PDMR2007} (accordingly, $S_{x_n}$, the spectral density of the measurement noise in \cite{PDMR2007} is two times larger than our $S_{x_n}$).}.

Similarly, for model P:
\begin{subequations}
\label{EqPSDP}
\begin{align}
\label{EqPSDPa}
S_{xx}(\omega)&=\left[ \frac{(1 +\tau^2\omega^2)/m^2}{\vert D(\omega)\vert^2}\right]S_{F_{\textnormal{th}}}+ \left[\frac{g^2\omega^2/\tau_0^2}{\vert D(\omega)\vert^2}\right]S_{x_n}
\\
\label{EqPSDPb}
S_{x'x'}(\omega)&=(1 +\tau^2\omega^2)S_{y,y}(\omega)\nonumber\\
&=\left[ \frac{ (1 +\tau^2\omega^2)/m^2}{\vert D(\omega)\vert^2}\right]S_{F_{\textnormal{th}}}+ \left[\frac{(1 +\tau^2\omega^2)[(\omega_0^2-\omega^2)^2+\omega^2/\tau_0^2]}{\vert D(\omega)\vert^2}\right] S_{x_n} \ .
\end{align}
\end{subequations}
%S_{xy}(\omega)&=\left[-\frac{(1+i\omega \tau)/m^2}{\vert D(\omega)\vert^2}\right]S_{F_{\textnormal{th}}}+\left[g (\omega/\tau_0)\frac{\omega/\tau_0+i(\omega_0^2-\omega^2)}{\vert D(\omega)\vert^2}\right] S_{x_n}
The only difference with the PSDs of model V is the absence  of the factor $\omega^2$ in the terms proportional to $S_{F_{\textnormal{th}}}$. 
Eqs. (\ref{EqPSDP}) reduce to Eqs. (3)  and (4) of  \cite{PDMR2007} for $\tau=0$. One can also check that $T_{\textnormal{eff}}^{(x)}\equiv k<x^2>_{\textnormal{st}}=1/(2\pi) \int_{-\infty}^{\infty} S_{xx}(\omega) d\omega$ and $T_{\textnormal{eff}}^{(v)}\equiv m<v^2>_{\textnormal{st}}=1/(2\pi) \int_{-\infty}^{\infty} S_{vv}(\omega) d\omega$, which shows that the divergence of the kinetic temperature  as $\tau \rightarrow 0$ is related to the behavior of $S_{vv}(\omega)=\omega^2 S_{xx}(\omega)$ at high frequencies.

\section{Logratio of the path probabilities in the NESS (model V)}

In this Appendix we compute the quantity $\ln {\cal P}_+[\{x_s\}]/{\cal P}_-[\{\hat x_s\}]$  in the NESS for model V. 

In general, the path probability ${\cal P}_+[\{x_s\}]$ can be obtained via two different routes. First, one can start from  the conditional probability  ${\cal P}[{\{\bf X}_s\}\vert {\bf X}_0]$ of the path ${\{\bf X}_s\}_{s\in[0,t]}=(\{x_s\},\{y_s\})_{s\in[0,t]}$ given by Eq. (\ref{EqProba}) in Appendix C and perform the  path integral over $\{y_s\}$. Second, one can consider the  Langevin equation with memory and colored noised obtained by inserting the integrated expression of $y(t)$ in Eq. (\ref{EqmodelVa}). This is the procedure used in \cite{CPV2012} for a very similar (but overdamped) model\footnote{As explained in \cite{CPV2012}, this route is only valid if the initial condition $y_0$ is chosen from a specific random distribution. Hence, this route is useful in the asymptotic long-time limit only.}.
 In the NESS, however, one can simply use the fact  that $ {\cal P}_+[\{x_s\}]$ is Gaussian and given in the Fourier (frequency) domain by
\begin{align}
\label{EqPomega}
{\cal P}_+[\{x_s\}]\propto \exp\Big[-\frac{1}{2}  \int_{-\infty}^{\infty} \frac{d\omega}{2\pi}  x(\omega)S_{xx}(\omega)^{-1} x(-\omega)\Big]
\end{align}
where $S_{xx}(\omega)=S_{vv}(\omega)/\omega^2$ is the power spectrum distribution of the position $x$ (the normalization factor does not play any role in what follows). We stress that the influence of the initial conditions is neglected when going to the frequency domain, which is correct as long as one only considers  expectation values\footnote{Indeed, we have shown in \cite{MR2012} that the so-called  `boundary' terms  may have a dramatic effect on  large fluctuations, a problem that occurs when the position and velocity of the particle are unbounded; this issue was already pointed out in \cite{ZBCK2005}. To obtain the correct expression of the boundary terms, there is no other choice than performing the functional integration of ${\cal P}[\{{\bf X}_s\}]$, which is a workable but tedious calculation.}.  Using the expression of $S_{vv}(\omega)$ given by Eq. (\ref{EqPSDVa}) and replacing $S_{v_n}$ by $2T'/\gamma'$, we obtain after some simple manipulations
\begin{align}
\label{EqPomega}
{\cal P}_+[\{x_s\}]\propto \exp\Big[-\frac{m^2}{4\gamma T}  \int_{-\infty}^{\infty} \frac{d\omega}{2\pi}  x(\omega) \frac{\vert D(\omega)\vert^2}{\frac{\gamma T +\gamma' T'}{\gamma T}+\omega_0^2 \tau^2} x(-\omega)\Big] 
\end{align}
where $D(\omega)$ is given by Eq. (\ref{EqDomega}). The probability ${\cal P}_-[\{\hat x_s\}]$ for the time-reversed trajectory is then obtained by replacing $x(\omega)$ by $x(-\omega)$ and changing  $\gamma'$ to $-\gamma'$. This readily yields 
\begin{align}
\ln \frac{{\cal P}_+[\{x_s\}]}{{\cal P}_-[\{{\hat x}_s\}]}=-\frac{\gamma'}{\gamma  T}\int_{-\infty}^{\infty} \frac{d\omega}{2\pi}  \: \omega^2 x(\omega) \frac{\gamma+k\tau-m\tau \omega^2}{\frac{\gamma T +\gamma' T'}{\gamma T} +\omega^2\tau^2}x(-\omega) \ .
\end{align}
This logratio depends on the measurement noise via the presence of $T'$ in the denominator. However, this dependence disappears when performing the average. Indeed, since $S_{xx}(\omega)\equiv <x(\omega)x(-\omega)>_{\textnormal{st}}$ by definition, we obtain
\begin{align}
\frac{1}{t}<\ln \frac{{\cal P}_+[\{x_s\}]}{{\cal P}_-[\{\hat {x}_s\}]}>_{\textnormal{st}}&= -\frac{\gamma'}{\gamma T}\int_{-\infty}^{\infty} \frac{d\omega}{2\pi}  \:  \frac{\gamma+k\tau-m\tau \omega^2}{ \frac{\gamma T +\gamma' T'}{\gamma T} +\omega^2\tau^2}S_{vv}(\omega)\nonumber\\
&= -\frac{2\gamma'}{m^2}\int_{0}^{\infty} \frac{d\omega}{\pi} \:  \frac{\omega^2(\gamma +k\tau -m\tau \omega^2) }{\vert D(\omega)\vert^2} \ .
\end{align}
This integral can be computed analytically and after some tedious but elementary algebra we obtain the very simple result 
\begin{align}
\label{EqratioP3}
\frac{1}{t}<\ln \frac{{\cal P}_+[\{x_s\}]}{{\cal P}_-[\{\hat {x}_s\}]}>_{\textnormal{st}}=\frac{ \gamma '^2}{m\gamma_{\textnormal{eff}}}
\end{align}
with $\gamma_{\textnormal{eff}}$ given by Eq. (\ref{Eqgtilde}). This result generalizes Eq. (\ref{Eqratio0V}) for $\tau >0$ but cannot be  taken as a pertinent definition of the EP rate since it is independent of the measurement noise.

\section{Entropy production in the super-system (model V)}

In this Appendix, we compute the entropy production in model V when the full statistical information on  the microscopic degrees of freedom is available, that is
when the two trajectories $\{x_s\}_{s\in[0,t]}$ and $\{y_s\}_{s\in[0,t]}$ can be observed. The EP along the trajectory ${\{\bf X}_s\}_{s\in[0,t]}$ can then be  defined in the standard way from the logratio of the probabilities of the forward and reverse paths.  Since the two noises $\xi(t)$ and $\eta(t)$ are independent and Gaussian distributed, the probability of the path ${\{\bf X}_s\}$, conditioned on the initial state ${\bf X}_0=(x_0,v_0,y_0)$, is given by
\begin{align}
\label{EqProba}
{\cal P}[{\{\bf X}_s\}\vert {\bf X}_0]&\propto \exp\left[-\frac{1}{4\gamma T}\int_0^t ds\: \Big(m \ddot x_s+\gamma \dot x_s+\gamma' y_s+kx_s\Big)^2 -\frac{1}{2S_{v_n}}\int_0^t ds\: \Big(\tau \dot y_s+y_s-\dot x_s\Big)^2\right] \ .
\end{align}
The backward path $\{ \hat {\bf X}_s\}_{s\in[0,t]}$ is then defined by the time-reversal operation $\hat x_s\equiv x_{t-s},\dot{\hat x}_s\equiv -\dot x_{t-s},\hat y_s\equiv y_{t-s},\hat{ \dot y}_s\equiv -\dot y_{t-s}$, which yields
\begin{align}
{\cal P}[\{ \hat {\bf X}_s\}\vert \hat {\bf X}_0]&\propto \exp\left[-\frac{1}{4\gamma T}\int_0^t ds\: (m \ddot x_s-\gamma \dot x_s+\gamma' y_s+k x_s)^2 -\frac{1}{2S_{v_n}}\int_0^t ds\: ( -\tau \dot y_s+y_s+\dot x_s)^2\right]  \ .
\end{align}
Hence
\begin{align}
\label{Eqratio}
\ln  \frac{{\cal P}[\{ {\bf X}_s\}\vert {\bf X}_0]}{{\cal P}[\{ \hat {\bf X}_s\}\vert \hat {\bf X}_0]}= \Delta s_m[\{ {\bf X}_s\}]+\Delta s_{m'}[\{ {\bf X}_s\}]
\end{align}
where 
\begin{subequations}
\begin{align}
\Delta s_{m}[{\{\bf X}_s\}]&=-\frac{1}{T} \int_0^t ds\: (m \ddot x_s+\gamma' y_s+k x_s)\circ \dot x_s\\
\Delta s_{m'}[{\{\bf X}_s\}]&=-\frac{2}{S_{v_n}}\int_0^t ds\: \left[\tau \dot y_s -\dot x_s\right] \circ y_s  \ .
\end{align}
\end{subequations}
The total entropy production along the trajectory is then given by
\begin{align}
\label{Eqeprxs}
\sigma[{\{\bf X}_s\}]\equiv \ln  \frac{{\cal P}[\{ {\bf X}_s\}]}{{\cal P}[\{ \hat {\bf X}_s\}]}=\Delta s_{\textnormal{sys}}+\Delta s_m[{\{\bf X}_s\}]+\Delta s_{m'}[{\{\bf X}_s\}] 
\end{align}
where $\Delta s_{\textnormal{sys}}=\ln p_0({\bf X}_0)/p_t({\bf X}_t)$. By construction, $\sigma[{\{\bf X}_s\}]$ satisfies the standard fluctuation theorems\cite{Sreview2012}. Note that $y$ has been treated as an even variable under time reversal in order to derive Eq. (\ref{Eqratio}) (despite the fact that it has the dimension of a velocity). This is indeed essential for recovering the correct expression of $\Delta s_{m}[{\{\bf X}_s\}]\equiv q[{\{\bf X}_s\}]/T$  where $q[{\{\bf X}_s\}]$ is the heat exchanged with the environment at temperature $T$, defined by Eq. (\ref{EqLheat}) in the main text\footnote{In fact, the correct expression of $\Delta s_{m}[{\{\bf X}_s\}]$ can also be recovered with $y$ odd and  $\gamma'$ changed to $-\gamma'$, but this does not give a sensible result for $\Delta s_{m'}[{\{\bf X}_s\}]$ since this quantity then vanishes for $\tau=0$. Here $y$ plays the same role as the auxiliary variables $v_i$ in \cite{PV2009} that appear when mapping a generalized Langevin equation with exponential memory kernel and colored noise to a set of coupled Markovian equations. These variables are indeed even under time reversal (the same is true for the variable $U$ defined by Eqs. (B1) in \cite{CPV2012}).}. 

If one ignores the contribution of the boundary terms  in Eq. (\ref{Eqeprxs}), which are not extensive in time and vanish on average in the NESS,  the entropy production for long times in given by
\begin{align}
\label{Eqeprlim}
\sigma[{\{\bf X}_s\}] \simeq \gamma' (\frac{1}{T'}-\frac{1}{T})\int_0^t ds\: y_s \circ \dot x_s  
\end{align}
where $T'=\gamma' S_{v_n}/2$ (this corresponds to Eq. (B7) in \cite{CPV2012}). 

At this stage, the quantity $\Delta s_{m'}[{\{\bf X}_s\}]$ has no definite physical meaning. However, if $T'$ is  the actual temperature of a second heat bath coupled to the Brownian particle, and Eq.  (\ref{EqmodelVb}) is rewritten as 
\begin{align}
(\gamma' \tau) \dot y +\gamma' (y-\dot x)=\sqrt{2\gamma' T'}\eta (t) \ , 
\end{align} 
then one can  identify $\Delta s_{m'}[{\{\bf X}_s\}]$ with $q'[{\{\bf X}_s\}]/T'$, where  
\begin{subequations}
\begin{align}
\label{EqLheatpa}
q'[{\{\bf X}_s\}]&=\int_0^t ds\:[\gamma' y_s-\sqrt{2\gamma T'}\eta_s]\circ y_s \\
\label{EqLheatpb}
&=\gamma' \int_0^t ds\: \left[-\tau \dot y_s +\dot x_s\right] \circ y_s  
\end{align}
\end{subequations}
is the heat exchanged with this second reservoir. Interestingly, Eq. (\ref{Eqeprlim}) is also the result obtained in \cite{K2012} for the EP associated to Eq. (\ref{EqLKQ2}) when the contribution of the feedback controller is included (indeed, note that the above equations have a well-defined limit for $\tau=0$). In this case, as noted in section 4A, $\Delta s_{m'}[{\{\bf X}_s\}]$ corresponds to the time-continuous limit\footnotemark[18] of the quantity $\Delta s_p$ which is interpreted in \cite{K2012} as the entropy production due to the measurement process. From the expression of $\Delta s_p$ (see Eq. (7) in \cite{K2012}), one can easily see that the definition of the (discrete) reverse process proposed in \cite{K2012}  amounts to treating $y$ as an even variable under time reversal in the continuous-time limit. In this respect, it is not surprising that  the EP of the full system  computed in \cite{K2012} obeys the detailed FT. 

Upon averaging, one recovers from Eq. (\ref{Eqeprxs}) the balance equation obtained from the time derivative of the Shannon entropy. The Fokker-Planck equation (\ref{EqFPV}) is then written as
\begin{align}
\label{EqFPsuper}
\partial_t  p_t({\bf X})=-\partial_x [vp_t({\bf X})] -\frac{1}{m}\partial_v[-(kx+\gamma' y) p_t({\bf X})+ J_t({\bf X})]-\frac{1}{\gamma' \tau}\partial_y [\gamma' vp_t({\bf X})+J_t'({\bf X})]
\end{align}
where $J_t({\bf X})$ and $J_t'({\bf X})$ are the irreversible components of the probability currents defined as
\begin{align}
\label{EqJ}
J_t({\bf X})&=-\gamma[v+\frac{T}{m}\partial_v \ln  p_t({\bf X})] p_t({\bf X}) \nonumber\\ 
J_t'({\bf X})&=-\gamma'[y+\frac{T'}{\gamma' \tau}\partial_y \ln  p_t({\bf X})] p_t({\bf X}) \ .
 \end{align}
(Note that $J_t'({\bf X})$ is indeed the time-antisymmetric component of $J_t^{(y)}=-\gamma'[(y-v)p_t({\bf X})+T'/(\gamma' \tau)\partial_y p_t({\bf X})] $  because $y$ is even under time reversal.) One then finds
\begin{align}
\label{Eq2ndL4}
\dot {\Sigma}(t)=\dot S_{\textnormal{sys}}(t)+\dot S_m(t)+\dot S_{m'}(t)
\end{align}
where  $\dot S_m(t)=-(1/T) \int d{\bf X} \: v \: J_t({\bf X})=-(1/T) \int dx dv \: v \: J_t(x,v)$ is again given by Eqs. (\ref{EqSm}), 
\begin{subequations}
\begin{align}
\label{EqSmxya}
\dot S_{m'}(t)&=-\frac{1}{T'} \int d{\bf X} \: y \: J_t'({\bf X})\\
\label{EqSmxyb}
&=\frac{\gamma'}{T'}[<y^2>_t-\frac{T'}{\gamma' \tau}] \ ,
\end{align}
\end{subequations}
and
\begin{align}
\dot {\Sigma}(t)=\frac{1}{\gamma T}\int d{\bf X} \: \frac{[ J_t({\bf X})]^2}{p_t({\bf X})}+\frac{1}{\gamma' T'}\int d{\bf X} \: \frac{[ J_t'({\bf X})]^2}{p_t({\bf X})} \ .
\end{align}
This latter expression is in agreement with the general expression of the EP rate in the full phase space when even and odd variables are present\cite{SF2012}. 
Furthermore, using the  inequality\footnote{This is the continuous version of the inequality given in \cite{VdBE2010} in the case of a discrete set of numbers (see also  \cite{V2012}).}
\begin{align}
 \int dy \: \frac{ [J_t({\bf X})]^2}{p_t({\bf X})} \ge  \frac{[\int dy \: J_t({\bf X})]^2}{\int dy \: p_t({\bf X})}=\frac{[J_t(x,v)]^2}{p_t(x,v)}\ ,
\end{align}
one finds that  
\begin{align}
\label{Eqinequality}
\dot {\Sigma}(t)\ge \dot {\tilde \Sigma}(t)+\frac{1}{\gamma' T'}\int d{\bf X} \: \frac{[ J_t'({\bf X})]^2}{p_t({\bf X})}  \ge \dot {\tilde \Sigma}(t)
\end{align}
where $\dot {\tilde\Sigma}(t)=1/(\gamma T)\int dx dv J_t(x,v)]^2/p_t(x,v)$ is the apparent EP rate considered in section 4.B. 

Finally, in the steady state, one obtains
 \begin{align}
\label{Eqsigmatotxy}
\dot \Sigma&=\dot Q(\frac{1}{T}-\frac{1}{T'})\nonumber\\
&=\frac{\gamma \gamma'}{m\gamma_{\textnormal{eff}}}\frac{(T-T')^2}{TT'} \ , 
\end{align}
which generalizes Eqs. (\ref{Eqsigmatot}) to $\tau >0$ as the heat flow $\dot Q$ is still given by Eqs. (\ref{EqSmSm'}) (with $T_{\textnormal{eff}}$ replaced by $T_{\textnormal{eff}}^{(v)}$).

\end{document}